\documentclass[aps,prb,twocolumn,amsmath,amssymb,superscriptaddress,floatfix,nofootinbib]{revtex4}

\usepackage{color}         
\usepackage{graphics}      
\usepackage[pdftex]{graphicx}      
\usepackage{bm}
\usepackage{natbib}            
\usepackage[colorlinks,plainpages=false,linkcolor=blue,urlcolor=blue,citecolor=blue,pdfpagemode=UseNone,pdfstartview=FitBH]{hyperref}
\usepackage{float}

\begin{document}

\title{Adiabatic variation of the charge-density-wave phase diagram\\in the 123 cuprate (Ca$_x$La$_{1-x}$)(Ba$_{1.75-x}$La$_{0.25+x}$)Cu$_3$O$_y$}

\author{M.~Bluschke}
\email[]{m.bluschke@fkf.mpg.de; keren@physics.technion.ac.il}
\affiliation{Max Planck Institute for Solid State Research, Heisenbergstr. 1, 70569 Stuttgart, Germany}
\affiliation{Helmholtz-Zentrum Berlin f\"{u}r Materialien und Energie, Wilhelm-Conrad-R\"{o}ntgen-Campus BESSY II, Albert-Einstein-Str. 15, 12489 Berlin, Germany}

\author{M.~Yaari}
\affiliation{Physics Department, Technion-Israel Institute of Technology, Haifa 32000, Israel}

\author{E.~Schierle}
\affiliation{Helmholtz-Zentrum Berlin f\"{u}r Materialien und Energie, Wilhelm-Conrad-R\"{o}ntgen-Campus BESSY II, Albert-Einstein-Str. 15, 12489 Berlin, Germany}

\author{G. Bazalitsky}
\affiliation{Physics Department, Technion-Israel Institute of Technology, Haifa 32000, Israel}

\author{J. Werner}
\affiliation{Institute for Solid State Research, Leibniz Institute for Solid State and Materials Research, Helmholtzstr. 20, 01069 Dresden, Germany}

\author{E.~Weschke}
\affiliation{Helmholtz-Zentrum Berlin f\"{u}r Materialien und Energie, Wilhelm-Conrad-R\"{o}ntgen-Campus BESSY II, Albert-Einstein-Str. 15, 12489 Berlin, Germany}

\author{A.~Keren}
\email[]{m.bluschke@fkf.mpg.de; keren@physics.technion.ac.il}
\affiliation{Physics Department, Technion-Israel Institute of Technology, Haifa 32000, Israel}

\begin{abstract}
In order to explain the emergence of the anomalous pseudogap state and high-temperature superconductivity in the cuprates, intense research activity over three decades has focused on unravelling the connection between the various instabilities of the underdoped regime. In the high-temperature superconductor (Ca$_x$La$_{1-x}$)(Ba$_{1.75-x}$La$_{0.25+x}$)Cu$_3$O$_y$ (CLBLCO) isovalent chemical substitution produces smooth changes to the CuO$_2$ plane buckling and the Cu(II)-to-apical-oxygen distance, allowing us to study the interdependence of charge-density-wave (CDW) order, superconductivity and the pseudogap at constant hole doping in two adiabatically connected representations of the 123 cuprate structure. In this study, resonant soft x-ray scattering measurements reveal the first observation of incommensurate CDW correlations in CLBLCO and demonstrate a lack of correlation between $T_{\text{CDW}}$ and the pseudogap crossover temperature ($T^{\ast}$). This result  disfavours a scenario in which the opening of the pseudogap at $T^{\ast}$ results from fluctuating CDW correlations.

\end{abstract}

\maketitle

\section{Introduction}

The underdoped regime of the cuprate superconductors is host to a complex set of nearly degenerate and strongly interdependent electronic, magnetic and structural ordering tendencies, whose relationships to one another and in particular to superconductivity have yet to be fully understood.\cite{Fradkin2015, Keimer2015} The pseudogap phase, which lies at the heart of the underdoped regime, is identified with the opening of a partial gap at the antinodal regions of the Fermi surface as observed by angle resolved photoemission spectroscopy as well as with a decrease of both the spin susceptibility and the specific heat.\cite{Lee2006} While some studies have discussed the reduced susceptibility below the pseudogap crossover temperature ($T^{\ast}$) in terms  of antiferromagnetic correlations,\cite{Alloul1989, YASUOKA1994} more recently evidence has been produced both for\cite{Fujita2014, Comin2014_Bi2201, Caprara2017} and against\cite{Santi_PRB_2014, Badoux2016, Peng2018} the notion that incommensurate charge-density-wave (CDW) order\cite{Ghiringhelli2012, Chang2012,  Santi_PRB_2014, Huecker2014, Comin2014_Bi2201, Neto_2014_Bi2212, Tabis_2014_Hg1201, Tabis2017} may result in the partial gap opening. Furthermore, various analytical studies based on the spin-fermion model have concluded that both $d$-wave superconductivity and charge order arise from a magnetically mediated interaction.\cite{Wang2014, Efetov2013, Montiel2017} In one case, fluctuations of an SU(2) matrix order parameter corresponding to the superposition of d-wave superconductivity and charge order was suggested to be responsible for the pseudogap behaviour.\cite{Efetov2013, Montiel2017}

\begin{figure*}[ht]
\includegraphics[width=16cm]{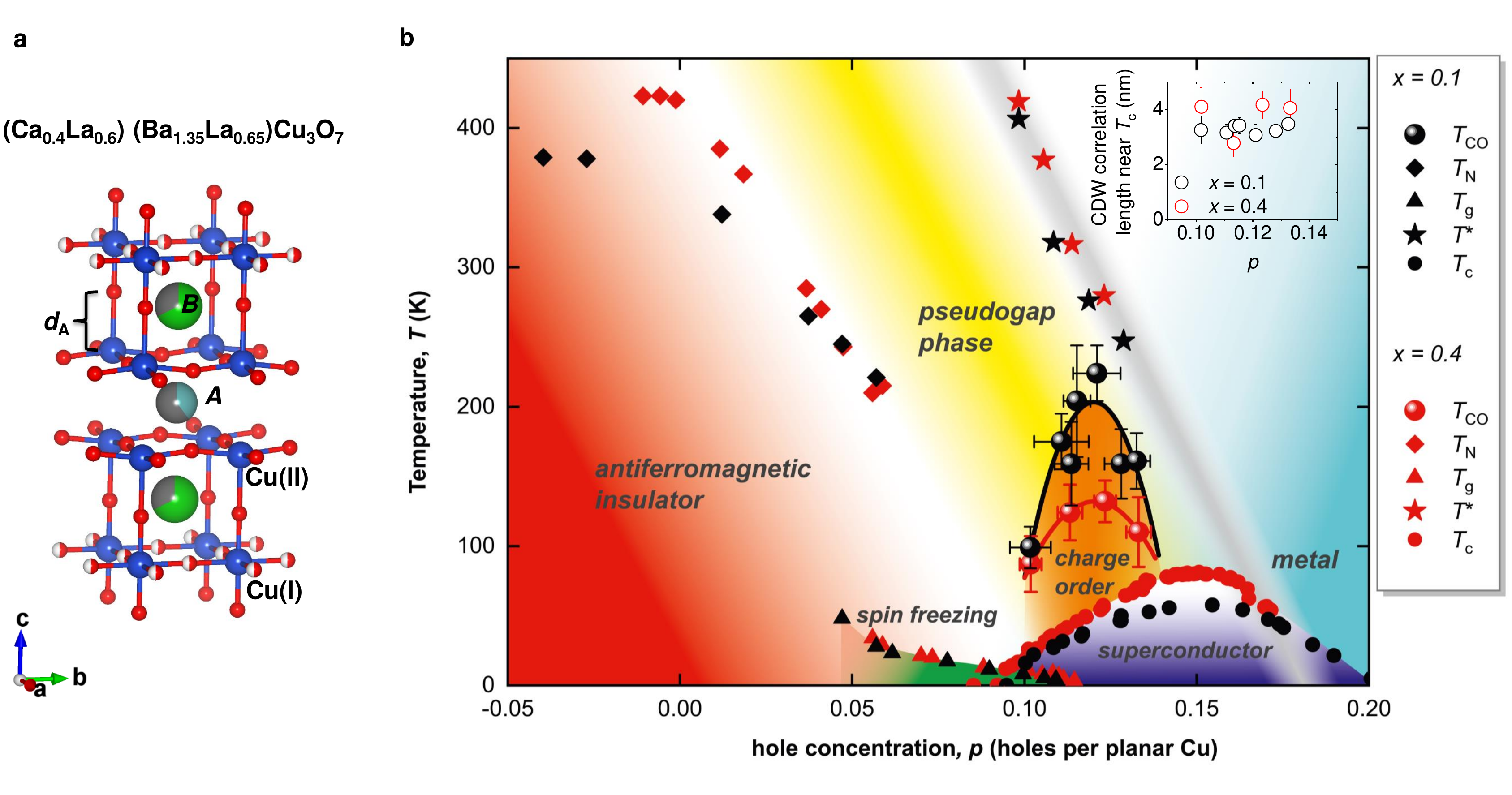}
\caption{
\textbf{ (Ca$_x$La$_{1-x}$)(Ba$_{1.75-x}$La$_{0.25+x}$)Cu$_3$O$_y$ structure and phase diagram.} (a) CLBLCO structure for $x=0.4$ and $y=7$. The partial occupation of the $A$ and $B$ sites by Ca, La and Ba are indicated by teal, grey and green colouring respectively. Similarly the half occupation of the oxygen sites in the charge reservoir layer are indicated by half-red/white spheres, as opposed to the fully occupied oxygen sites (fully red spheres) in the CuO$_2$ planes and at the apical sites. (b) $T_{\text{CDW}}$ measured in this study for single crystals with $x=0.1$ and $x=0.4$ are compared with $T_{\text{c}}$ extracted from resistivity measurements,\cite{Kanigel2002_PRL} $T^{\ast}$ from magnetization measurements\cite{Lubashevsky_PRB_2008} which are in agreement with measurements of the Knight shift,\cite{Cvitanic2014} and $T_{\text{g}}$ and $T_{\text{N}}$ both obtained from muon spin resonance.\cite{Kanigel2002_PRL, Ofer2006_magnetic_analog} The solid black and red curves are parabolic fits to the $T_{\text{CDW}}$ data sets for the $x=0.1$ and $x=0.4$ families respectively. The uncertainties in $p$ were extracted from the widths of the Meissner transitions used to characterize $T_{\text{c}}$ whereas the uncertainties in the onset temperatures were estimated from a sensitivity analysis of $T_{\text{CDW}}$ to the background subtraction procedure and to the number of points used in the linear fits to the onset of RXS intensity. The inset shows the CDW correlation length near $T_{\text{c}}$ as a function of doping for both families.
\label{fig:phase_diagram}}
\end{figure*}

The relationships between the various instabilities of the underdoped regime are difficult to assess, in part due to the specific properties of the different cuprate families exhibiting this phenomenology. Moving from one family to the next a large number of structural and chemical parameters are changed, resulting in a complex variation of the strength and characteristics of the ordering tendencies present in this doping range. For example, while the incommensurability of spin and charge stripe order in the underdoped La-based cuprates, such as La$_{2-x}$Sr$_x$CuO$_4$, are coupled to one another,\cite{Yamada_PRB_1998} this has proven to be a special case in contrast to the more general result observed in Y-, Hg- and Bi- based cuprates where the CDW incommensurability is decoupled from that of the incommensurate spin fluctuations and the CDW wavevector decreases with doping rather than increasing.\cite{Comin2014_Bi2201, Santi_PRB_2014, Tabis2017} Furthermore, recent studies of CDW order in the La-based cuprates La$_{1.475}$Nd$_{0.4}$Sr$_{0.125}$CuO$_4$, La$_{1.875}$Ba$_{0.125}$CuO$_4$ and La$_{1.65}$Eu$_{0.2}$Sr$_{0.15}$CuO$_4$ have highlighted the importance of structural degrees of freedom in determining the material specific charge order phenomenology, and in particular the coupling between CDW and proximal ordering tendencies within the underdoped regime, such as symmetry breaking structural distortions, nematicity\cite{Achkar2016} and spin order.\cite{Miao2017}

In order to understand the intrinsic relationships between the intertwined phases of the underdoped regime, one requires an adiabatic tuning parameter to manipulate and study their interdependence. By adiabatic tuning parameter we designate a degree of freedom of the system which can be manipulated to smoothly transform the electronic structure, without crossing bands. Here we control a coupled set of structural degrees of freedom, including the CuO$_2$ plane buckling angle $\Theta_{\text{b}}$ (defined as the deviation from a straight 180$^{\circ}$ Cu(II)-O-Cu(II) bond) and the Cu(II)-to-apical-oxygen distance $d_{\text{A}}$, both of which have been shown to be relevant in determining the strength of the planar superexchange coupling $J$.\cite{Pavarini2001, Petit2009, Peng2017, Bogdanov2018} In order to actively control $\Theta_{\text{b}}$ and $d_{\text{A}}$ while simultaneously minimizing variations in the crystal symmetry or chemical and magnetic degrees of freedom, we have prepared a series of (Ca$_x$La$_{1-x}$)(Ba$_{1.75-x}$La$_{0.25+x}$)Cu$_3$O$_{y}$ (CLBLCO) single crystals with varying oxygen content $y$ and varying ratio of Ca:Ba contents. The crystal structure of CLBLCO (depicted in Fig.~\ref{fig:phase_diagram}a) is of the $AB_2$Cu$_3$O$_{7}$ type similar to YBa$_2$Cu$_3$O$_{7-\delta}$ (YBCO), however in CLBLCO the La$^{3+}$ ions on the $B$ site bond oxygens from the charge-reservoir layer along both the $a$ and $b$ directions, disrupting the formation of CuO chain ordering and resulting in an average tetragonal structure.\cite{Goldschmidt1993} While increasing oxygen content introduces holes into the CuO$_2$ planes, allowing access to the entire underdoped regime, isovalent substitution of Ca for Ba shifts the La$^{3+}$ ions from the $A$ site to the $B$ site producing opposing changes in the average valence states and ionic radii of the $A$ and $B$ site cations. The redistribution of chemical pressure within the unit cell when increasing the Ca:Ba ratio has been shown to decrease $\Theta_{\text{b}}$ by up to $30\%$ and increase $d_{\text{A}}$ on the order of $1\%$.\cite{Chmaissem1999, Ofer2008} In addition, the overall lattice parameters are slightly modified, with the Cu(II)-O-Cu(II) bond length varying slightly on the order of about $0.1\%$. Previous studies have shown that the decrease of $\Theta_{\text{b}}$ and elongation of $d_{\text{A}}$ with increasing $x$ result in an enhanced superexchange coupling $J$\cite{Ellis2015_RIXS} and accordingly elevated N\'{e}el and spin freezing temperatures\cite{Kanigel2002_PRL, Ofer2006_magnetic_analog} ($T_{\text{N}}$ and $T_{\text{g}}$ in Fig.~\ref{fig:phase_diagram}b), as well as a slight modification of the hole doping efficiency (holes transferred to the CuO$_2$ plane per dopant oxygen). Here we report the first observation of CDW correlations in CLBLCO and, by studying the temperature-doping phase diagram of CDW order for two extreme values of the Ca:Ba ratio, we demonstrate that the energy scale governing the pseudogap is independent of the energy scale associated with CDW correlations in this system.

\section{Methods}
\subsection{crystal growth, preparation, and characterization}
\begin{figure}[ht]
\includegraphics[width=8.6cm]{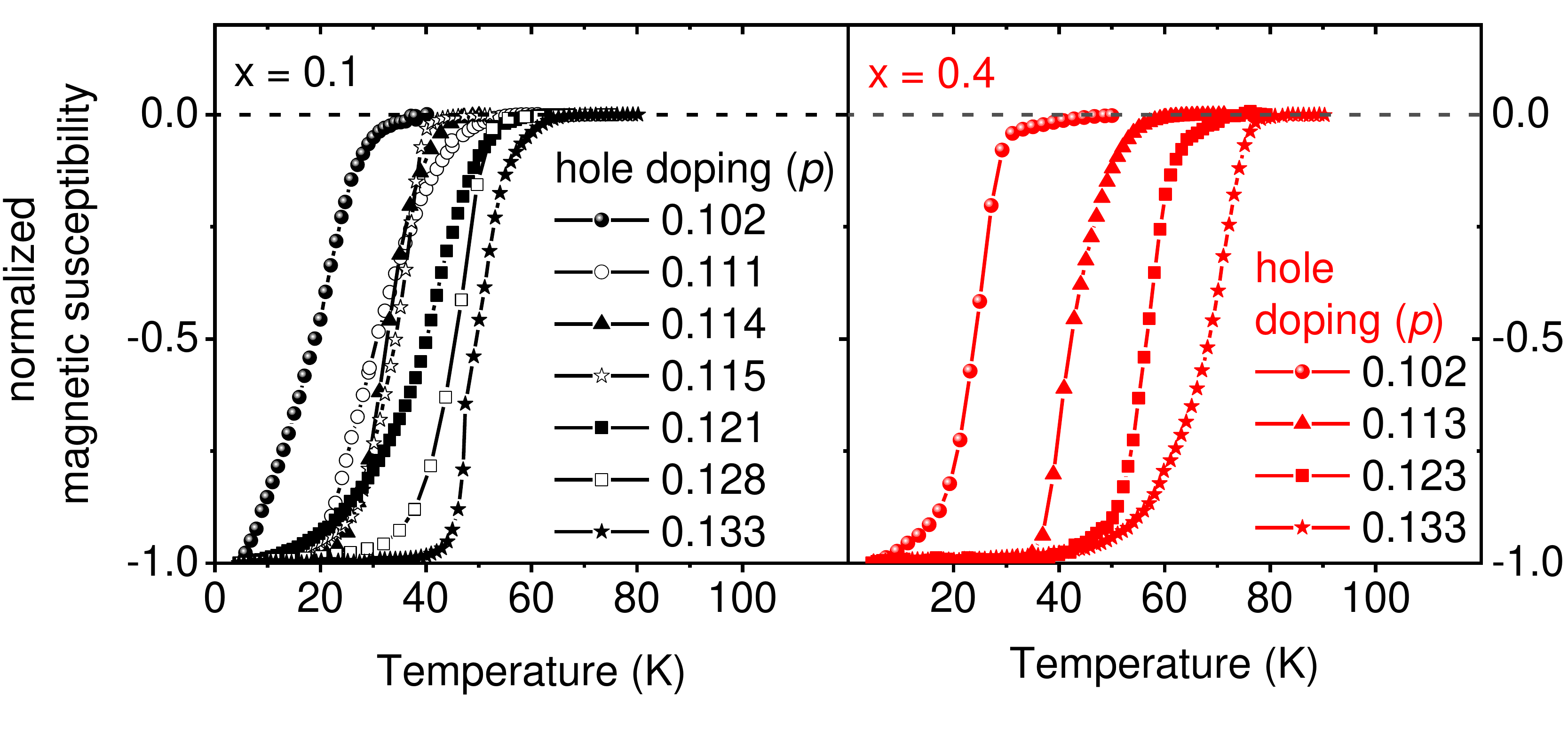}
\caption{\textbf{Magnetization.} Normalized magnetic susceptibility for all samples in this study. The superconducting critical temperature $T_{\text{c}}$ was taken to be the midpoint of the diamagnetic response and the uncertainty in the derived hole doping ($p$) was extracted from the width of the transition.
\label{fig:magnetization}}
\end{figure}

Two sets of (Ca$_x$La$_{1-x}$)(Ba$_{1.75-x}$La$_{0.25+x}$)Cu$_3$O$_y$ (CLBLCO) single crystals with $x=0.1$ and $x=0.4$ were grown with a floating zone furnace according to the recipe described in Ref. \onlinecite{Drachuck2012} and after preparing the powder according to Ref. \onlinecite{Goldschmidt1993}.  Following growth the crystals were annealed in either ambient pressure or a high pressure of 500 bars and at various temperatures in the vicinity of 500 $^{\circ}$C for 10 to 14 days in order to achieve oxygen homogeneity. The superconducting critical temperatures $T_{\text{c}}$ were determined via measurements of the Meissner effect using a SQUID magnetometer. The normalized magnetic susceptibility is plotted for each sample in Fig.~\ref{fig:magnetization}. The oxygen content $y$ was determined from $T_{\text{c}}$ according to the phase diagram in Ref. \onlinecite{Drachuck2012} and then converted to units of hole doping per planar Cu ($p$) according to the linear relation $p=K(y-y_{\text{N}})$ with $K=0.74$ and $0.45$, and $y_{\text{N}}=6.915$ and $6.78$, for the $x=0.1$ and $x=0.4$ families respectively. $K$ represents the doping efficiency and $y_{\text{N}}$ the oxygen content where $T_{\text{N}}$ begins to drop. The ratio between $K$'s for the two families has been previously determined by nuclear quadrupole resonance studies of CLBLCO powders\cite{Eran2010} and the scaling of $K$ in our study was chosen so that $T^{\text{max}}_{\text{c}}$ occurs at $p=0.15$. While the parameters for the linear transformation between $y$ and $p$ are only chosen to align the drop in $T_{\text{N}}$ and to align the maximum $T_{\text{c}}$ of the $x=0.1$ and $x=0.4$ families, this choice of parameters also has the effect that all characteristic doping levels of the phase diagram are the same for the two families. By characteristic doping levels we mean the $p$ where: $T_{\text{N}}$ starts to drop, antiferromagnetic long-range order is lost, superconductivity begins, the spin-glass phase ends, $T_{\text{c}}$ reaches its maximum value, and the superconducting dome ends. The fact that the same linear relation between $y$ and $p$ also transforms the beginnings and ends of the CDW domes (Fig. \ref{fig:phase_diagram}b) of the two families on to each other is a strong indication that our transformation is appropriate in contrast to a suggestion by Tallon.\cite{TallonPRB14} All samples in this study were oriented along the (001) crystallographic direction, and the in-plane orientations were determined via Laue diffraction prior to the soft x-ray scattering experiments. Typical crystal sizes were on the order of 1 mm x 1 mm x 0.1 mm.

\subsection{Resonant x-ray scattering}

The measured intensity in a resonant x-ray scattering (RXS) experiment is a function of the incident photon energy $\hbar\omega$ and can be expressed as\cite{Haverkort2010}

\begin{equation}
I(\omega) \propto \bigg| \sum_i e^{\imath\mathbf{q}\cdot\mathbf{r}_i} \bf\epsilon_{out}^* \cdot \textit{F}_i(\omega) \cdot \epsilon_{in}\bigg|^2 \; ,
\label{scattering_intensity}
\end{equation}
where the sum is over all atoms in the sample, $\bf{\epsilon}_{in}$ and $\bf{\epsilon}_{out}$ are the polarizations of the incoming and outgoing light, and $\mathbf{r}_i$ and $F_i(\omega)$ are the position and scattering tensor of the i$^{th}$ atom. The detector used in our experiments distinguishes neither the energy nor the polarization of the scattered photon, thereby yielding a signal which represents an integration over elastic and inelastic as well as polarization-conserving and polarization-rotating scattering processes. The element specific scattering tensor $F_i(\omega)$ has a non-trivial energy dependence which results from the unique multiplet structure of the interatomic electronic transitions of each ion. When tuning the photon energy to a particular resonance of $F_i(\omega)$, one not only enhances the scattering from ions of type $i$ with respect to distinct ions in the material under illumination, but one actually enhances the scattering from the specific electronic states involved in the core-to-valence excitation with respect to all other occupied and unoccupied electronic states of that ion.\cite{Fink_REXS_review} The Cu $L_3$ resonance designates the promotion of a 2$p$ core electron to the partially occupied 3$d$ shell. As such, Cu $L_3$ RXS provides information about spatial periodicities of the Cu 3$d$ valence states. Furthermore, the CLBLCO structure contains two distinct Cu sites with distinct valence states.\cite{Chmaissem_PRB_2001} The valence of the Cu(I) sites varies strongly as dopant oxygens are incorporated into the charge reservoir layer. In contrast the Cu(II) sites in the CuO$_2$ planes receive only a moderate hole doping and are characterized roughly by a 3$d^9$ electronic configuration. Throughout this study the incident photon energy was kept constant at 931.5~eV in order to optimize sensitivity to charge superstructures in the CuO$_2$ planes. Likewise the polarization of the incident light was set perpendicular to the scattering plane such that $\epsilon_{in}$ remains parallel to the CuO$_2$ planes for all angles of incidence.

\begin{figure}[ht]
\includegraphics[width=8.6cm]{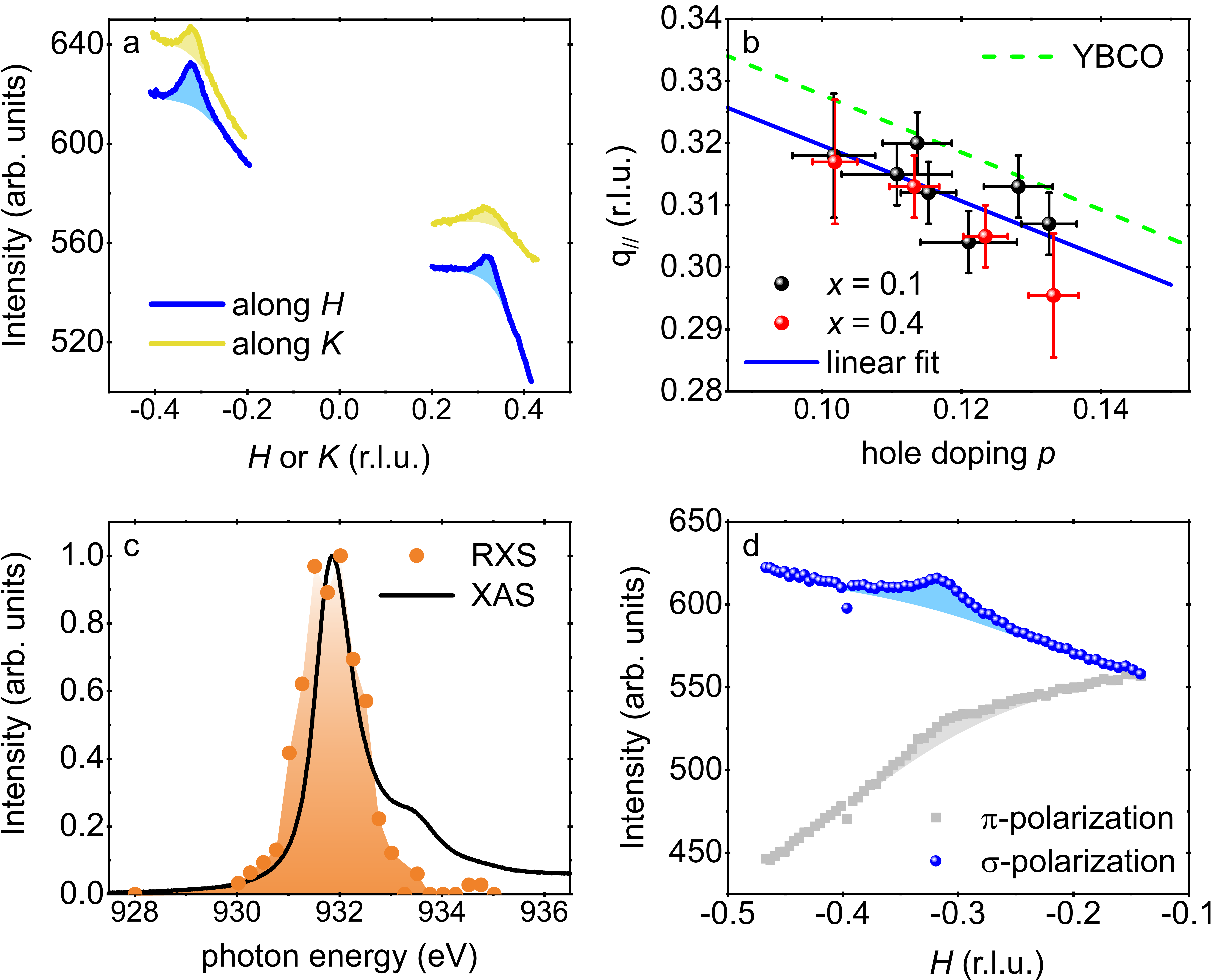}
\caption{\textbf{Characterization of the CDW reflection.} (a) Example of raw data taken at 12 K for the sample with $x=0.1$ and $p = 0.115$. The CDW peak is clearly observed for positive and negative in-plane momentum transfer and is equivalent along both in-plane directions of the tetragonal unit cell. (b) Doping dependence of the in-plane momentum transfer $q_{//}$ of the CDW reflection. Data points from both Ca contents $x=0.1$ and $x=0.4$ are plotted together and fit with a line. The dashed green line represents a linear fit to the $q_{//}$ measured in YBCO.\cite{Santi_PRB_2014} The uncertainties in $p$ and $q_{//}$ were extracted respectively from the width of the Meissner transition and from the sensitivity of $q_{//}$ to the RXS background subtraction procedure. (c) Energy dependence of the CDW peak in CLBLCO $x = 0.1$,  $p = 0.114$ compared with the x-ray absorption spectrum measured on the same crystal. (d) Raw scan through the CDW reflection at positive $H$, for varying incident photon polarization.
\label{fig:energy_scattering_geometry}}
\end{figure}

The resonant x-ray scattering experiments were performed using the ultra-high-vacuum-compatible x-ray ultraviolet diffractometer\cite{Fink_REXS_review} of the UE46-PGM1 beamline\cite{ENGLISCH_2001_UE46} at the BESSY II synchrotron in Berlin. The x-ray ultraviolet diffractometer is equipped with two motorized circles. All measurements of the charge order reflections reported in this study were performed in a geometry as close to back-scattering as allowed by the finite extension of the detector ($2\theta \sim 165^{\circ}$). With the detector angle fixed, the CDW peak was probed via rocking scans (varying the angle of the sample with respect to the beam). Equivalent scans at high temperature were used to determine the shape of the background, and small offsets between the high and low temperature data were removed by fitting a line to the end points of the residuals. Projecting the scans onto the reciprocal space direction $H$ the background-subtracted data were fit with a Lorentzian function in order to extract the peak area and correlation length according to $\xi_{//} = \frac{a}{\pi \times \text{FWHM}}$. The samples were mounted with the CuO$_2$ planes perpendicular to the scattering plane, and slight deviations from this condition were corrected for by scanning the detector position perpendicular to the horizontal scattering plane. Where possible, a manual \textit{in-situ} rotation about the (001) direction was used to align the (-101) Bragg reflection, i.e. align the in-plane Cu-O bond directions parallel and perpendicular to the scattering plane respectively. 

\begin{figure*}[ht]
\includegraphics[width=18cm]{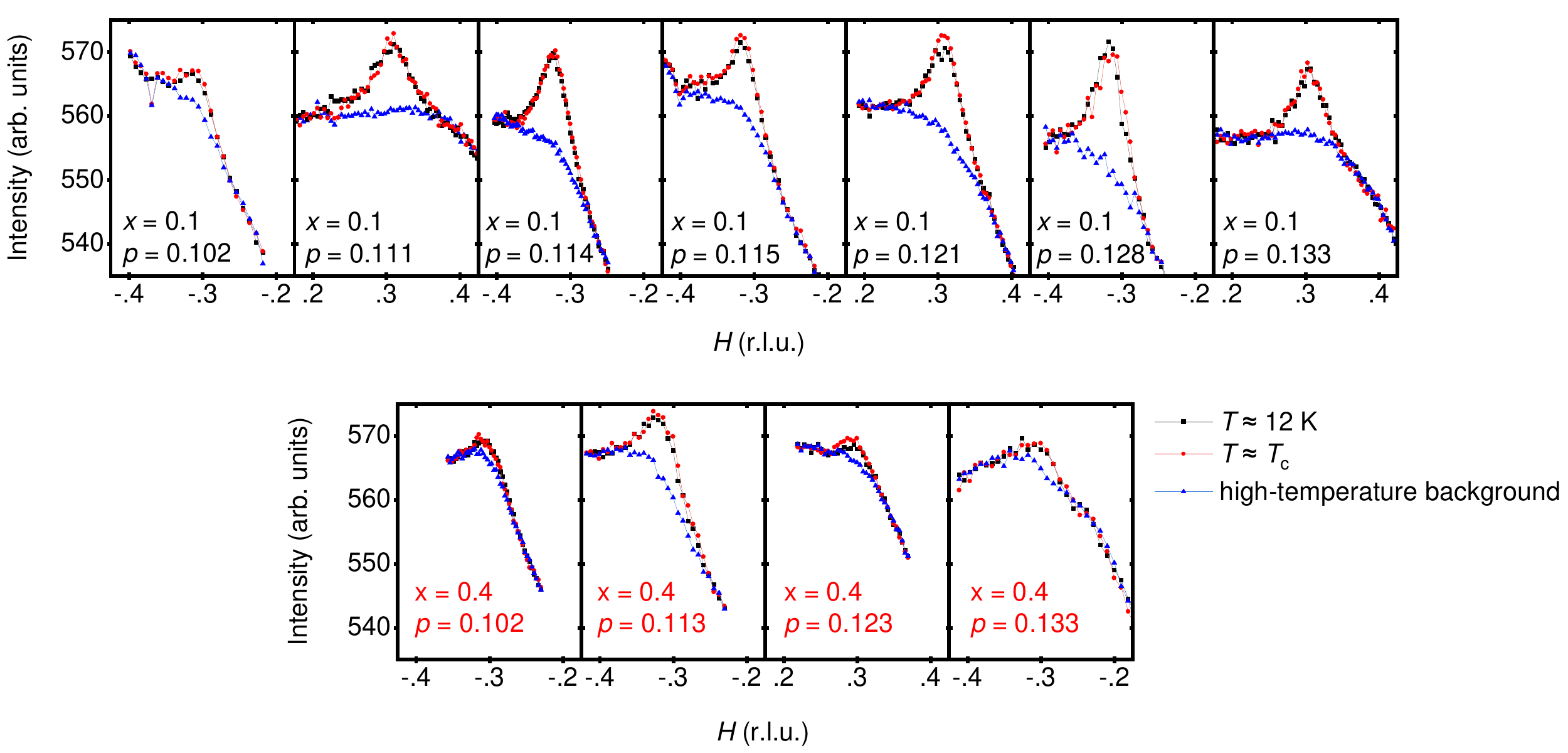}
\caption{
\textbf{Reciprocal space scans.} Rocking scans (projected onto $H$) through the CDW peak for various doping levels $p$ in the $x=0.1$ and $x=0.4$ families. A linear offset has been applied to match the background intensity of the low- and high-temperature scans. The intensity scale is identical for each plot and offsets on the order $\pm 100$ have been applied to each data set to bring the scans into the field of view. }  \label{fig:triple_panel}
\end{figure*}

\section{Results}

\subsection{Incommensurate CDW on a tetragonal lattice}

Resonant x-ray scattering measurements were performed on two families of CLBLCO single crystals, where each family is described by a fixed Ca:Ba ratio parameterized by the corresponding values $x=0.1$ and $x=0.4$. For each family a series of single crystals were prepared with varying oxygen contents, allowing access to a doping range between approximately $p=0.1$ and $p=0.14$. Incommensurate CDW correlations with real space periodicities ranging between approximately $3.1$$a$ and $3.4$$a$ were detected along both in-plane Cu-O-Cu bond directions. Consistent with the reported tetragonality of the CLBLCO structure,\cite{Goldschmidt1993} the component of the CDW wavevector within the CuO$_2$ plane, $q_{//}$, is found to be identical along both in-plane directions. Fig.~\ref{fig:energy_scattering_geometry}a shows raw data from   (Ca$_{0.1}$La$_{0.9}$)(Ba$_{1.65}$La$_{0.35}$)Cu$_3$O$_{7.03}$ indicating the presence of weak diffraction peaks positioned symmetrically for positive and negative in-plane momentum transfer at $q_{//} = \pm0.312$ reciprocal lattice units (r.l.u.). The CDW wavevector was determined for each sample by measuring the reflection for both positive and negative in-plane momentum transfer, and averaging the extracted peak positions. $q_{//}$ was found to decrease with increasing hole doping consistent with trends observed in YBCO as well as both Bi- and Hg-based cuprates.\cite{Comin2014_Bi2201, Santi_PRB_2014, Tabis2017} The measured values of $q_{//}$ for the $x=0.1$ and $x=0.4$ families are plotted together in Fig. \ref{fig:energy_scattering_geometry}b along with a linear fit to the combined data sets and the linear trend extracted in the same manner from measurements performed on single crystals of YBCO.\cite{Santi_PRB_2014}

\subsection{Cu $L_3$ resonance and dichroism}
 
Due to the relatively small number of charges involved in the CDW superstructure, the associated soft x-ray scattering cross-section is extremely weak and difficult to detect on the background of charge scattering originating from all electrons of all the ions in the material. Accordingly the CDW reflections studied here are only detectable for incident photons tuned precisely to the Cu $L_3$ resonance of the participating Cu ions. Fig.~\ref{fig:energy_scattering_geometry}c presents the integrated peak intensity as a function of incident photon energy for $\sigma$-polarized photons and a (Ca$_{0.1}$La$_{0.9}$)(Ba$_{1.65}$La$_{0.35}$)Cu$_3$O$_{7.02}$ sample. The CDW reflection is strongest for the photon energy 931.5~eV, approximately aligned with the main maximum of the x-ray absorption spectrum. This resonance is primarily associated with transitions to valence states of the Cu(II) ions, indicating that the incommensurate CDW correlations exist predominantly in the CuO$_2$ planes.\cite{Achkar_PRL_2012_Distinct_Charge_Orders} The CDW reflection is observed to be most intense for incident $\sigma$-polarized photons~(Fig.~\ref{fig:energy_scattering_geometry}d), reflecting the strong in- vs. out-of-plane linear dichroism of the x-ray absorption at the Cu $L_3$ edge.

\subsection{Temperature-doping phase diagram}

The strength of the CDW correlations observed in underdoped CLBLCO varies non-monotonically as a function of hole doping $p$, with the strongest diffraction intensities and highest onset temperatures ($T_{\text{CDW}}$) obtained near $p = 0.12$. In Fig. \ref{fig:phase_diagram}b $T_{\text{CDW}}$ is plotted as a function of $p$ for $x=0.1$ and $x=0.4$. Although we cannot strictly claim the absence of CDW correlations for $p < 0.10 $ and  $p > 0.14$, we were not able to identify experimental signatures of CDW order for these dopings. The full set of samples in which we were able to identify and characterize CDW reflections is listed in Table~\ref{tab:fit_params} along with their corresponding $x$, $y$, $p$, and $T_{\text{c}}$, as well as the $T_{\text{CDW}}$, $q_{//}$ and the in-plane correlation lengths $\xi_{//}$ extracted from the measurements.

\begin{figure*}[ht]
\includegraphics[width=18cm]{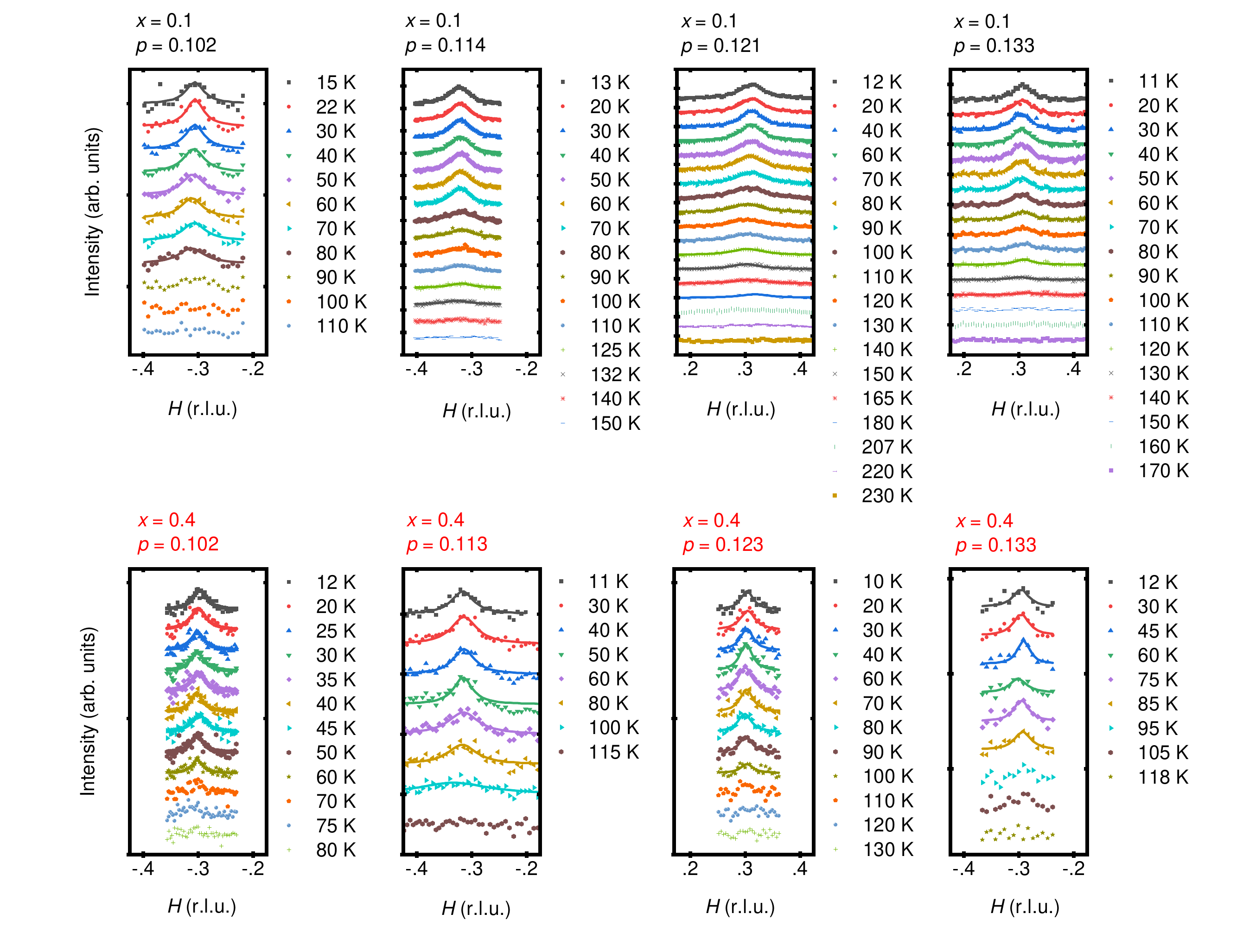}
\caption{
\textbf{Temperature series and Lorentzian fitting.} The CDW peak is presented for pairs of samples with similar dopings from the $x=0.1$ and $x=0.4$ families. A high-temperature background (as in Fig.~\ref{fig:triple_panel}) has been subtracted from each scan. Wherever possible, a Lorentzian fit was performed to the background subtracted data. An individual offset has been applied to each scan for clarity. The separation between tick marks on the intensity axes corresponds to the same absolute intensity difference in all plots.
\label{fig:8_samples_H-scans_sub_bkgd}}
\end{figure*}

As a function of temperature the CDW reflection onsets gradually demonstrating no well-defined transition. Fig. \ref{fig:triple_panel} shows rocking scans through the CDW peak measured at base temperature and near $T_{\text{c}}$ along with the high-temperature background for all samples in which CDW order was identified. Four pairs of samples with similar dopings from the $x=0.1$ and $x=0.4$ families are compared in greater detail in Figs.~\ref{fig:8_samples_H-scans_sub_bkgd}, \ref{fig:T-dep_all_samples} and \ref{fig:T-dep_all_samples_correlation-length}. Fig.~\ref{fig:8_samples_H-scans_sub_bkgd} shows the background-subtracted data for each temperature together with the corresponding Lorentzian fit. An offset is applied to each scan for clarity. The integrated peak area after subtraction of the high-temperature background is plotted in Fig. \ref{fig:T-dep_all_samples} and the correlation length extracted from the Lorentzian fits is shown in Fig. \ref{fig:T-dep_all_samples_correlation-length}.  We note that the temperature at which CDW order is first observed may depend on the intensity of the reflection and the sensitivity of the detector, thereby leading to an underestimation of the onset temperature. Accordingly, a linear fit to the temperature-dependent peak area was performed near the onset of CDW order, and $T_{\text{CDW}}$ was assigned to the zero of this fit. In this way, the approximately linear onset of CDW intensity is extrapolated into the temperature regime where CDW diffraction intensity is hidden in the noise of the measurement. Despite the inherent difficulty in assigning a value to the onset temperature of CDW order, care was taken to measure all samples in the same manner such that the extracted values of $T_{\text{CDW}}$ can be considered as an approximate measure of the intrinsic energy scale governing the CDW correlations.

Naturally, one might expect that the increased chemical disorder on the $A$ and $B$ sites, which follows from increasing $x$, would promote the pinning of fluctuating CDW correlations. In contrast however, for lower $x$ the CDW order demonstrates a much stronger RXS intensity and has a higher associated energy scale as evidenced by the comparison of the $T_{\text{CDW}}(p)$ trends for $x=0.1$ and $x=0.4$ presented in Fig.~\ref{fig:phase_diagram}b.
We interpret the stronger RXS intensity in the $x=0.1$ family as resulting from a greater number of charges involved in the modulation, either in the sense of volume fraction or modulation amplitude. On the other hand, the correlation length in the $x=0.4$ family appears to be larger than in the  $x=0.1$ family where the onset temperature is highest. This unexpected behaviour may point towards the role of a structural  degree of freedom in determining the CDW correlation length. The disorder potential associated with dopant oxygens was suggested to be responsible for the finite CDW correlation length in YBa$_2$Cu$_3$O$_{7-\delta}$~\cite{Caplan_PRB_2015} and HgBa$_2$CuO$_{4+\delta}$~\cite{Tabis2017}, as well as crucial in understanding the crossover from 2D to 3D correlations in applied magnetic fields.\cite{Caplan_PRL_2017} In CLBLCO, the La$^{3+}$ ions on the $B$ site are responsible for disrupting the formation of CuO chain order in the charge reservoir layer. Clusters or formations of $B$-site La$^{3+}$ ions and the surrounding chain-layer oxygens may produce a disorder potential which limits the extent of the CDW phase coherence. The lack of a significant doping dependence of $\xi_{//}$ (inset of Fig.~\ref{fig:phase_diagram}b) is consistent with the suggestion that $\xi_{//}$ may be determined by an $x$ dependent disorder potential.

In the $x=0.1$ family the intensity of the CDW signal is suppressed upon cooling through $T_{\text{c}}$, evidencing a competition between the CDW and superconducting phases, consistent with the results of experiments performed on YBa$_2$Cu$_3$O$_{7-\delta}$.\cite{Santi_PRB_2014} Similarly, the in-plane correlation length $\xi_{//}$ increases with decreasing temperature exhibiting a plateau or slight suppression below $T_{\text{c}}$. In contrast, the competition between CDW and superconductivity is somewhat reduced for the $x=0.4$ family, especially with respect to the correlation length (Fig.~\ref{fig:T-dep_all_samples_correlation-length}). This behaviour is similar to that observed in cuprates such as Bi$_2$Sr$_{2-x}$La$_x$CuO$_{6+\delta}$,\cite{Comin2014_Bi2201} electron-doped Nd$_{2-x}$Ce$_x$CuO$_4$\cite{Neto_SciAdv_2015} and HgBa$_2$CuO$_{4+\delta}$\cite{Tabis2017} where CDW correlations were observed to be significantly weaker than in YBCO, and little or no competition was observed. By capturing both strong and weak competition in two smoothly connectable cuprate structures (CLBLCO $x=0.1$ and $x=0.4$), we demonstrate that the reduced competition in the three former compounds may be an inherent feature of weaker CDW correlations rather than some compound specific phenomenology. Possibly for the same reason, the competition between CDW and superconductivity in YBCO is most pronounced in the doping range where CDW order is strongest.\cite{Santi_PRB_2014}

\section{Discussion \& Outlook}

\subsection{Relation to the pseudogap}

The relationship of the pseudogap phase to charge ordering is debated. While some studies have suggested that the zero-field incommensurate charge peaks observed in RXS are not related to the opening of the pseudogap\cite{DallaTorre_2015_NJP, DallaTorre_2016_PRB}, many models have proposed an intimate relationship between the CDW and pseudogap phenomena,\cite{Berg_2009_NJP, Efetov2013, Lee2014, Wang2014, Fradkin2015, Caprara2017, Montiel2017} for example in which the pseudogap is thought of as a precursor phase to CDW order or is directly associated with CDW fluctuations. Experimental evidence for correlation between the doping dependent $T^{\ast}$ and $T_{\text{CDW}}$ was obtained from Knight shift and RXS measurements in Bi$_2$Sr$_{2-x}$La$_x$CuO$_{6+\delta}$,\cite{Comin2014_Bi2201} and then called into question after the discovery of a charge order modulation deep in the overdoped regime of (Bi,Pb)$_{2.12}$Sr$_{1.88}$CuO$_{6+\delta}$ detectable for temperatures well above $T^{\ast}$.\cite{Peng2018} In the 123 cuprate YBCO, the nonmonotonic doping dependence of $T_{\text{CDW}}$ compared to the monotonic increase of $T^{\ast}$ with decreasing $p$ led to the conclusion that the CDW observed in RXS is not responsible for the opening of the pseudogap.\cite{Santi_PRB_2014} However, the suppression of CDW correlations as doping is decreased towards $p\sim0.08$ may simply result from proximity to the incommensurate magnetically ordered ground state with which CDW order is known to compete,\cite{Santi_PRL_2013} or from CDW phase fluctuations which are enhanced by electronic correlations as the Mott insulating phase is approached.\cite{Caprara2017} We note that in CLBLCO the onset of spin-freezing with decreasing doping coincides with the maximum in $T_{\text{CDW}}$ (Fig. \ref{fig:phase_diagram}b), hinting at a phase competition scenario. In YBCO, on the higher doping side of the CDW dome $T^{\ast}$ and $T_{\text{CDW}}$ loosely track one another with the condition $T_{\text{CDW}} < T^{\ast}$ being satisfied at all dopings and $p_{\text{CDW}} < p^{\ast}$ at zero temperature\cite{Badoux2016} (where $p_{\text{CDW}}$ and $p^{\ast}$ are the critical dopings where CDW induced Fermi surface reconstruction and the pseudogap phase end respectively). The authors of Ref.~\onlinecite{Caprara2017} argue that while static CDW order is responsible for the Fermi surface reconstruction with electron pockets, CDW fluctuations extend to higher dopings and temperatures resulting in the electronic crossovers observed at $p^{\ast}$ and $T^{\ast}$. To test whether CDW correlations could be responsible for the opening of the pseudogap, without the complications of doping dependent phase competition and enhanced quantum fluctuations near critical dopings, we have exploited the ability to manipulate the structural and electronic properties of CLBLCO while remaining at constant doping. In the present study we have demonstrated a clear lack of positive correlation between the CDW and pseudogap energy scales upon varying $\Theta_{\text{b}}$ and $d_{\text{A}}$ at constant doping (on both sides of the CDW dome). Based on this observation, we argue that the opening of the pseudogap at $T^{\ast}$ does not result from the charge ordering instability. On the other hand, given that the condition $T_{\text{CDW}} < T^{\ast}$ holds for all dopings where CDW order was observed, we do not rule out the possibility that in CLBLCO the CDW may be an instability exclusively of the pseudogap regime.

\begin{figure}[ht]
\includegraphics[width=8.4cm]{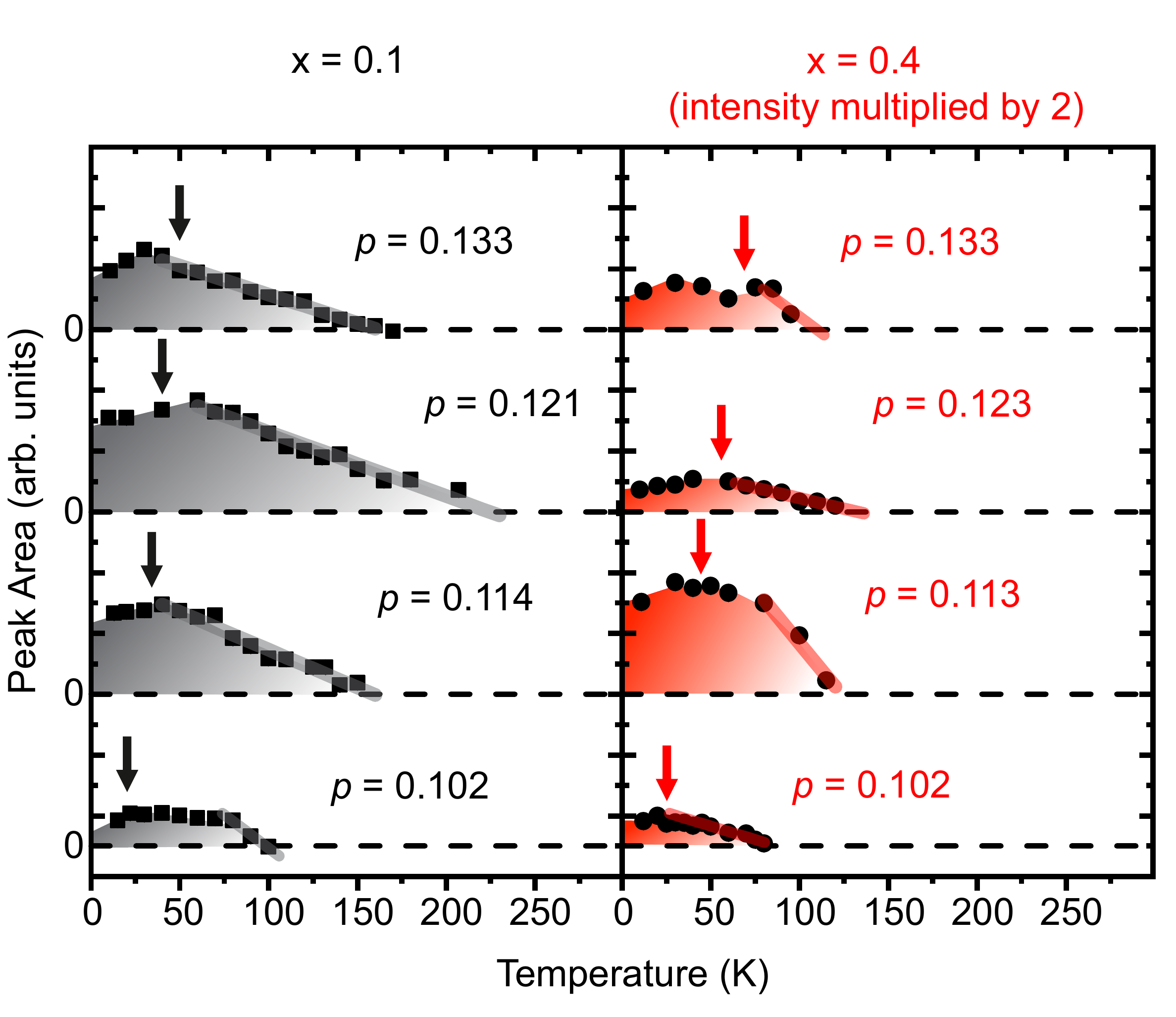}
\caption{\textbf{CDW peak area.} Temperature dependence of the integrated peak area for pairs of crystals from the $x=0.1$ and $x=0.4$ families with similar dopings. All data sets were measured upon warming and analyzed by subtracting a high-temperature background from the low temperature scattering data before integrating. The $T_{\text{CDW}}$ plotted in Fig.~\ref{fig:phase_diagram}b and listed in Table~\ref{tab:fit_params} were extracted from linear fits to the onset of scattering intensity as shown here. The arrows indicate $T_{\text{c}}$ for each sample. 
\label{fig:T-dep_all_samples}}
\end{figure}

\subsection{Relation to antiferromagnetism, superconductivity and preformed pairs}

Both theoretical and experimental results have supported the notion of a coupling between CDW order and antiferromagnetic fluctuations.\cite{Sachdev2013, Efetov2013, Wang2014, Neto2018} Various theoretical works\cite{Metlitski2010, Efetov2013} identified an emergent low-energy SU(2) symmetry relating d-wave superconductivity (particle-particle pairing) to charge ordering (particle-hole pairing), and the combined order parameter was shown to arise in the vicinity of an antiferromagnetic quantum critical point.\cite{Efetov2013} However, in our study the strongest CDW correlations are observed in the family with the weakest antiferromagnetism immediately calling into question whether CDW order could arise from antiferromagnetic quantum critical fluctuations. Rather, our results are consistent with recent quantum Monte Carlo simulations, which concluded that while superconductivity is enhanced in the vicinity of an antiferromagnetic quantum critical point\cite{Gerlach2017, Wang_Wang_Schattner_Berg_Fernandes_PRL_2018}, CDW order is not universally associated with antiferromagnetic quantum criticality.\cite{Wang_Wang_Schattner_Berg_Fernandes_PRL_2018}

\begin{figure}[ht]
\includegraphics[width=8.4cm]{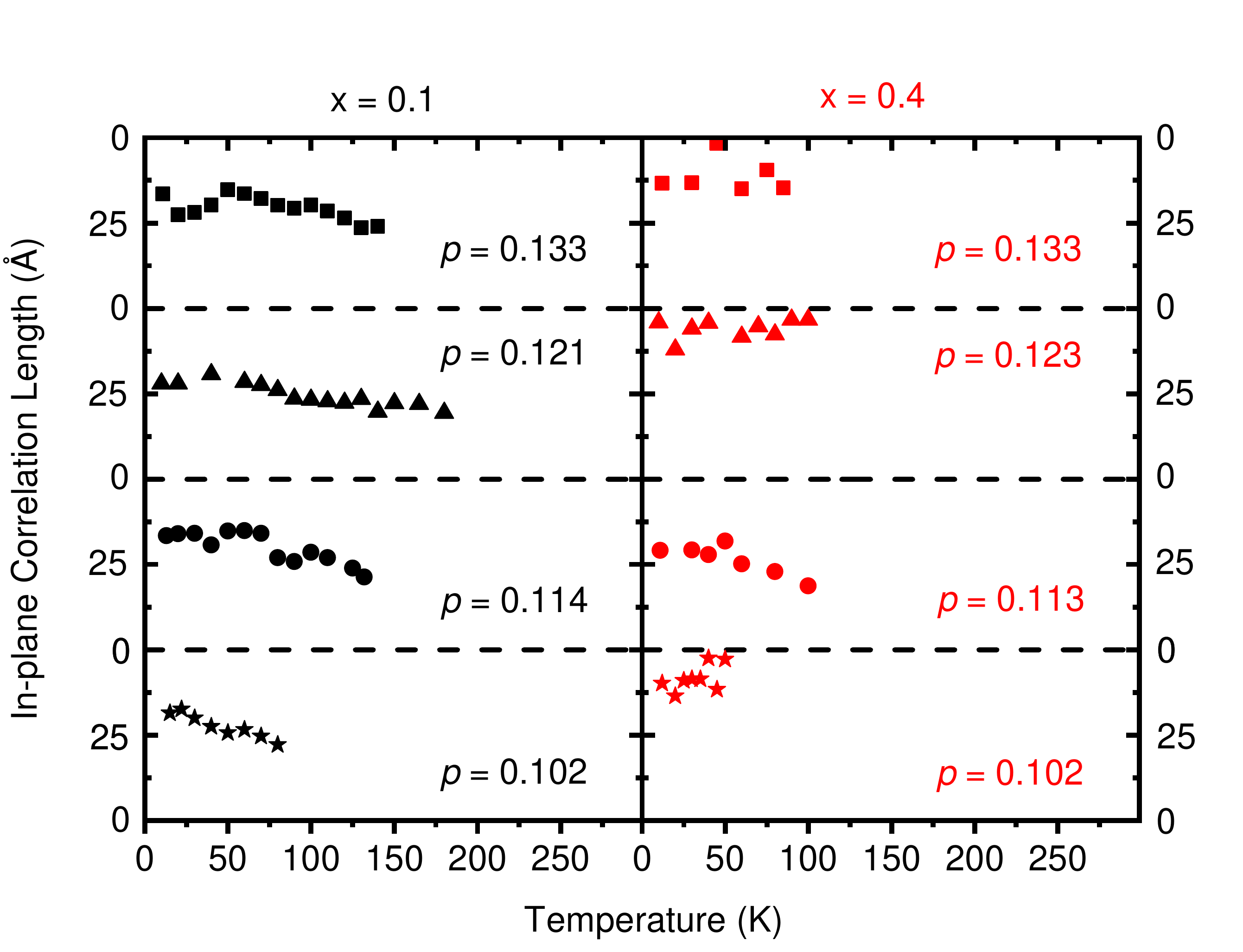}
\caption{
\textbf{CDW correlation length.} Temperature dependence of the CDW correlation length measured in pairs of crystals from the $x=0.1$ and $x=0.4$ families with similar dopings. All data sets were measured upon warming. The correlation lengths are extracted from Lorentzian fits to the background-subtracted data according to $\xi_{//} = \frac{a}{\pi \times \text{FWHM}}$. 
\label{fig:T-dep_all_samples_correlation-length}}
\end{figure}

The slight suppression of the CDW intensity and correlation length when cooling below $T_{\text{c}}$ ($x=0.1$ in Figs.~\ref{fig:T-dep_all_samples} and~\ref{fig:T-dep_all_samples_correlation-length}) indicates a competition between CDW order and superconductivity in CLBLCO. Additionally, we have shown that the combination of increasing $\Theta_{\text{b}}$ and decreasing $d_{\text{A}}$ simultaneously suppresses superconductivity and strengthens CDW correlations, similar to the effects of magnetic field and uniaxial pressure in YBCO.\cite{Santi_PRB_2014, Kim2019} The opposing responses of CDW and superconductivity when moving between families may represent a new manifestation of competition between the two phases. However, due to the elevated CDW onset temperatures ($T_{\text{CDW}} > T_{\text{c}}$) this would imply a competition with phase-incoherent Cooper-pairing above $T_{\text{c}}$, supporting a recent prediction of competition between the two order parameters in the fluctuating regime.\cite{Wang_Wang_Schattner_Berg_Fernandes_PRL_2018} Furthermore, $T_{\text{c}}$ continues to demonstrate a marked sensitivity to Ca:Ba substitution in optimally and overdoped CLBLCO indicating that while CDW order may be suppressed by strong superconducting fluctuations, superconductivity itself responds for all dopings to some other $x$-dependent parameter. Alternately, CDW order and superconductivity may respond independently to the varying Ca:Ba ratio by drawing on, or returning charge carriers to, the surrounding bath of metallic electrons. However, a complete decoupling of the two phases is considered unlikely in view of the wealth of experimental evidence for competition in both CLBLCO and underdoped cuprates in general.

\begin{table*}
	\begin{tabular}{  p{1.7cm}   p{1.7cm}  p{3cm}  p{1.9cm}  p{1.9cm}  p{2.9cm}  p{2.8cm}  }
		\hline \hline

		\hspace{0.05cm} x & \hspace{0.05cm} y & \hspace{0.05cm} hole doping $p$ \hspace{0.15cm}  &   $T_{\text{c}}$ (K) \hspace{0.15cm} &  $T_{\text{CDW}}$ (K) \hspace{0.15cm} & in-plane correlation length near $T_{\text{c}}$ (nm) \hspace{0.15cm} &  \hspace{0.5cm}  $q_{//}$ (r.l.u.) \\

		\hline \hline
		\hspace{0.05cm} 0.1 & \hspace{0.05cm} 6.99 & \hspace{0.05cm} 0.102(6) &  19(6) & 99(15)  & 3.3(5) & \hspace{0.5cm}  0.318(10) \\
		\hspace{0.05cm} 0.1 & \hspace{0.05cm} 7.02 & \hspace{0.05cm} 0.111(8) &  31(8) & 175(20)  & 3.2(3) & \hspace{0.5cm}  0.315(5) \\
		\hspace{0.05cm} 0.1 & \hspace{0.05cm} 7.02 &  \hspace{0.05cm} 0.114(5) &  33(5) & 159(30) &  3.4(4)  & \hspace{0.5cm} 0.320(5) \\
		\hspace{0.05cm} 0.1 & \hspace{0.05cm} 7.03 & \hspace{0.05cm} 0.115(4) &   34(4) & 204(40) &  3.4(3) & \hspace{0.5cm}  0.312(5) \\
		\hspace{0.05cm} 0.1 & \hspace{0.05cm} 7.05 & \hspace{0.05cm} 0.121(7) &   40(7) & 224(20) &  3.1(4) &  \hspace{0.5cm} 0.304(5)\\
		\hspace{0.05cm} 0.1 & \hspace{0.05cm} 7.07 & \hspace{0.05cm}  0.128(5) &   46(5) & 159(25) & 3.2(4) & \hspace{0.5cm} 0.313(5)  \\
		\hspace{0.05cm} 0.1 & \hspace{0.05cm} 7.08 & \hspace{0.05cm}  0.133(4) &   50(4) &  161(20) & 3.5(4) & \hspace{0.5cm} 0.307(5)  \\
		\hline \hline		
		\hspace{0.05cm} 0.4 & \hspace{0.05cm} 6.92 &  \hspace{0.05cm} 0.102(3) &   24(4)  &  87(20)  &  4.1(7) &  \hspace{0.5cm} 0.308(10) \\
		\hspace{0.05cm} 0.4 & \hspace{0.05cm} 6.98 &  \hspace{0.05cm} 0.113(4) &   43(5)  &  124(20)  &  2.8(5) &  \hspace{0.5cm} 0.313(5) \\
		\hspace{0.05cm} 0.4 & \hspace{0.05cm} 7.03 &  \hspace{0.05cm} 0.123(3) &   57(4)  &  132(15)  &  4.2(5) &  \hspace{0.5cm} 0.305(5) \\
		\hspace{0.05cm} 0.4 & \hspace{0.05cm} 7.07 &  \hspace{0.05cm} 0.133(4) &   69(5)  &  110(25)  &  4.1(7) &  \hspace{0.5cm} 0.295(10) \\
		\hline \hline
	\end{tabular}
	\caption{\textbf{Lorentzian fitting of CDW peaks.} After subtraction of a high-temperature background the in-plane correlation length $\xi_{//}$ and the in-plane wavevector $q_{//}$ are extracted from Lorentzian fits to the CDW peaks and tabulated for all of the crystals in this study. $T_{\text{CDW}}$ is taken to be the zero of a linear fit to the temperature dependent integrated peak area. The crystals are identified by the Ca content $x$, as well as the oxygen content $y$ and hole doping $p$, of which the last two are derived from the superconducting critical temperature $T_{\text{c}}$ as determined from the Meissner transition. }
	\label{tab:fit_params}
\end{table*}

\subsection{Relation to Fermi-surface reconstruction}
Measurements of the Hall, and the Seebeck coefficients\cite{LeBoeuf2007, Chang2010, Doiron-Leyraud_2013_Hg1201} as well as quantum oscillations\cite{Doiron-Leyraud2007, Yelland2008, Barisic2013} in YBa$_2$Cu$_3$O$_{7-\delta}$, YBa$_2$Cu$_4$O$_{8}$, and HgBa$_2$CuO$_{4+\delta}$ have demonstrated a generic reconstruction of the cuprate Fermi surface in the underdoped regime at low temperature and high magnetic fields. This Fermi surface reconstruction is associated with the formation of an electron pocket which brings about a sign change of the Hall resistivity, and is believed to be caused by the translational symmetry breaking CDW order. In addition to 2D CDW correlations observed at zero-field, nuclear magnetic resonance\cite{Wu2011} and high-field x-ray scattering experiments\cite{Gerber949, Chang2016} observed the presence of long-range 3D CDW correlations in magnetic fields exceeding 15~T. Recently, high-magnetic-field sound velocity measurements in YBCO\cite{Laliberte2018} have succeeded in tracking the onset fields and temperatures of 3D CDW order across a wide range of dopings and clearly distinguishing them from the onset of Fermi surface reconstruction observed in the Hall resistivity. The authors of Ref. \onlinecite{Laliberte2018} conclude therefore that the 3D CDW order  has minimal effect on the low-energy states of underdoped YBCO, and argue that the biaxial 2D CDW order is responsible for the formation of the electron pocket observed in high magnetic fields. In order to test this argument we propose CLBLCO as an ideal system for probing the correlation between 2D CDW order and Fermi surface reconstruction. Due to the significant variation in $T_{\text{CDW}}$  between the CLBLCO $x=0.1$ and $x=0.4$ families, we suggest that if 2D CDW order were responsible for the observed Fermi surface reconstruction, then a comparative transport study of the two families should reveal a related change in the temperature scale associated with Fermi surface reconstruction.

\section{Conclusion}
In summary, resonant soft x-ray scattering measurements have been used to obtain the first evidence for 2D CDW correlations in the bulk superconductor (Ca$_x$La$_{1-x}$)(Ba$_{1.75-x}$La$_{0.25+x}$)Cu$_3$O$_y$. By varying the oxygen content $y$, CDW correlations were detected within the doping range $0.102 < p < 0.133$. As a function of the chemical substitution $x$, which varies $\Theta_{\text{b}}$ and $d_{\text{A}}$ at approximately constant hole doping, we report a dramatic response of $T_{\text{CDW}}$ in contrast to the lack of a corresponding variation in $T^{\ast}$. These observations pose a challenge to recent theoretical models in which the pseudogap opens as a result of incommensurate CDW fluctuations,\cite{Caprara2017} or fluctuations of a combined CDW and superconducting order parameter.\cite{Efetov2013, Montiel2017} Furthermore, the opposing responses of $T_{\text{CDW}}$ and the superexchange coupling $J$ to the structural changes which depend on $x$,  appear to be at odds with suggestions of a magnetic origin of the charge ordering instability\cite{Efetov2013, Wang2014}. Looking forward, the significant variation in the onset temperature of zero-field (2-dimensional) CDW order in CLBLCO when comparing the $x=0.1$ and $x=0.4$ families may be exploited in future transport experiments to test whether 2-dimensional CDW order, rather than 3-dimensional CDW order, is responsible for the Fermi surface reconstruction observed in underdoped cuprates. More generally, we demonstrate that isovalent chemical substitution to produce adiabatically connected representations of a single structure type is a powerful method to study the cuprate phase diagram as a function of fundamental electronic parameters (like the superexchange $J$). This method provides a strategy for observing the interdependence of intertwined phases, while avoiding the complications of doping dependent phase competitions and enhanced fluctuations near critical dopings.

\section{Acknowledgements}
The authors thank B. Keimer, M. Minola, M. Le Tacon, D. Orgad, E. Dalla Torre, E. da Silva Neto, W. Tabis and T. Loew for fruitful discussions and critical reading of the manuscript. We thank the German Science Foundation (DFG) for financial support under grant No. SFB/TRR 80.


\begin{thebibliography}{67}%
\makeatletter
\providecommand \@ifxundefined [1]{%
 \@ifx{#1\undefined}
}%
\providecommand \@ifnum [1]{%
 \ifnum #1\expandafter \@firstoftwo
 \else \expandafter \@secondoftwo
 \fi
}%
\providecommand \@ifx [1]{%
 \ifx #1\expandafter \@firstoftwo
 \else \expandafter \@secondoftwo
 \fi
}%
\providecommand \natexlab [1]{#1}%
\providecommand \enquote  [1]{``#1''}%
\providecommand \bibnamefont  [1]{#1}%
\providecommand \bibfnamefont [1]{#1}%
\providecommand \citenamefont [1]{#1}%
\providecommand \href@noop [0]{\@secondoftwo}%
\providecommand \href[0]{\begingroup \@sanitize@url \@href}%
\providecommand \@href[1]{\@@startlink{#1}\@@href}%
\providecommand \@@href[1]{\endgroup#1\@@endlink}%
\providecommand \@sanitize@url [0]{\catcode `\\12\catcode `\$12\catcode
  `\&12\catcode `\#12\catcode `\^12\catcode `\_12\catcode `\%12\relax}%
\providecommand \@@startlink[1]{}%
\providecommand \@@endlink[0]{}%
\providecommand \url  [0]{\begingroup\@sanitize@url \@url }%
\providecommand \@url [1]{\endgroup\@href {#1}{\urlprefix }}%
\providecommand \urlprefix  [0]{URL }%
\providecommand \Eprint [0]{\href}%
\providecommand \doibase [0]{http://dx.doi.org}%
\providecommand \selectlanguage [0]{\@gobble}%
\providecommand \bibinfo  [0]{\@secondoftwo}%
\providecommand \bibfield  [0]{\@secondoftwo}%
\providecommand \translation [1]{[#1]}%
\providecommand \BibitemOpen [0]{}%
\providecommand \bibitemStop [0]{}%
\providecommand \bibitemNoStop [0]{.\EOS\space}%
\providecommand \EOS [0]{\spacefactor3000\relax}%
\providecommand \BibitemShut  [1]{\csname bibitem#1\endcsname}%
\let\auto@bib@innerbib\@empty
\bibitem [{\citenamefont {Fradkin}\ \emph {\textit{et~al.}}(2015)\citenamefont
  {Fradkin}, \citenamefont {Kivelson},\ and\ \citenamefont
  {Tranquada}}]{Fradkin2015}%
  \BibitemOpen
  \bibfield  {author} {\bibinfo {author} {\bibfnamefont {E.}~\bibnamefont
  {Fradkin}}, \bibinfo {author} {\bibfnamefont {S.~A.}\ \bibnamefont
  {Kivelson}}, \ and\ \bibinfo {author} {\bibfnamefont {J.~M.}\ \bibnamefont
  {Tranquada}},\ }\href{\doibase/10.1103/RevModPhys.87.457} {\bibfield
  {journal} {\bibinfo  {journal} {Rev. Mod. Phys.}\ }\textbf {\bibinfo {volume}
  {87}},\ \bibinfo {pages} {457} (\bibinfo {year} {2015})}\BibitemShut
  {NoStop}%
\bibitem [{\citenamefont {Keimer}\ \emph {\textit{et~al.}}(2015)\citenamefont
  {Keimer}, \citenamefont {Kivelson}, \citenamefont {Norman}, \citenamefont
  {Uchida},\ and\ \citenamefont {Zaanen}}]{Keimer2015}%
  \BibitemOpen
  \bibfield  {author} {\bibinfo {author} {\bibfnamefont {B.}~\bibnamefont
  {Keimer}}, \bibinfo {author} {\bibfnamefont {S.~A.}\ \bibnamefont
  {Kivelson}}, \bibinfo {author} {\bibfnamefont {M.~R.}\ \bibnamefont
  {Norman}}, \bibinfo {author} {\bibfnamefont {S.}~\bibnamefont {Uchida}}, \
  and\ \bibinfo {author} {\bibfnamefont {J.}~\bibnamefont {Zaanen}},\
  }\href{https://doi.org/10.1038/nature14165} {\bibfield  {journal} {\bibinfo
  {journal} {Nature}\ }\textbf {\bibinfo {volume} {518}},\ \bibinfo {pages}
  {179} (\bibinfo {year} {2015})}\BibitemShut {NoStop}%
\bibitem [{\citenamefont {Lee}\ \emph {\textit{et~al.}}(2006)\citenamefont
  {Lee}, \citenamefont {Nagaosa},\ and\ \citenamefont {Wen}}]{Lee2006}%
  \BibitemOpen
  \bibfield  {author} {\bibinfo {author} {\bibfnamefont {P.~A.}\ \bibnamefont
  {Lee}}, \bibinfo {author} {\bibfnamefont {N.}~\bibnamefont {Nagaosa}}, \ and\
  \bibinfo {author} {\bibfnamefont {X.-G.}\ \bibnamefont {Wen}},\
  }\href{\doibase/10.1103/RevModPhys.78.17} {\bibfield  {journal} {\bibinfo
  {journal} {Rev. Mod. Phys.}\ }\textbf {\bibinfo {volume} {78}},\ \bibinfo
  {pages} {17} (\bibinfo {year} {2006})}\BibitemShut {NoStop}%
\bibitem [{\citenamefont {Alloul}\ \emph {\textit{et~al.}}(1989)\citenamefont
  {Alloul}, \citenamefont {Ohno},\ and\ \citenamefont {Mendels}}]{Alloul1989}%
  \BibitemOpen
  \bibfield  {author} {\bibinfo {author} {\bibfnamefont {H.}~\bibnamefont
  {Alloul}}, \bibinfo {author} {\bibfnamefont {T.}~\bibnamefont {Ohno}}, \ and\
  \bibinfo {author} {\bibfnamefont {P.}~\bibnamefont {Mendels}},\
  }\href{\doibase/10.1103/PhysRevLett.63.1700} {\bibfield  {journal} {\bibinfo
  {journal} {Phys. Rev. Lett.}\ }\textbf {\bibinfo {volume} {63}},\ \bibinfo
  {pages} {1700} (\bibinfo {year} {1989})}\BibitemShut {NoStop}%
\bibitem [{\citenamefont {Yasuoka}\ \emph {\textit{et~al.}}(1994)\citenamefont
  {Yasuoka}, \citenamefont {Kambe}, \citenamefont {Itoh},\ and\ \citenamefont
  {Machi}}]{YASUOKA1994}%
  \BibitemOpen
  \bibfield  {author} {\bibinfo {author} {\bibfnamefont {H.}~\bibnamefont
  {Yasuoka}}, \bibinfo {author} {\bibfnamefont {S.}~\bibnamefont {Kambe}},
  \bibinfo {author} {\bibfnamefont {Y.}~\bibnamefont {Itoh}}, \ and\ \bibinfo
  {author} {\bibfnamefont {T.}~\bibnamefont {Machi}},\
  }\href{\doibase/10.1016/0921-4526(94)91811-2} {\bibfield  {journal} {\bibinfo
   {journal} {Physica B}\ }\textbf {\bibinfo {volume} {199-200}},\ \bibinfo
  {pages} {278} (\bibinfo {year} {1994})}\BibitemShut {NoStop}%
\bibitem [{\citenamefont {Fujita}\ \emph {\textit{et~al.}}(2014)\citenamefont
  {Fujita}, \citenamefont {Kim}, \citenamefont {Lee}, \citenamefont {Lee},
  \citenamefont {Hamidian}, \citenamefont {Firmo}, \citenamefont
  {Mukhopadhyay}, \citenamefont {Eisaki}, \citenamefont {Uchida}, \citenamefont
  {Lawler}, \citenamefont {Kim},\ and\ \citenamefont {Davis}}]{Fujita2014}%
  \BibitemOpen
  \bibfield  {author} {\bibinfo {author} {\bibfnamefont {K.}~\bibnamefont
  {Fujita}}, \bibinfo {author} {\bibfnamefont {C.~K.}\ \bibnamefont {Kim}},
  \bibinfo {author} {\bibfnamefont {I.}~\bibnamefont {Lee}}, \bibinfo {author}
  {\bibfnamefont {J.}~\bibnamefont {Lee}}, \bibinfo {author} {\bibfnamefont
  {M.~H.}\ \bibnamefont {Hamidian}}, \bibinfo {author} {\bibfnamefont {I.~A.}\
  \bibnamefont {Firmo}}, \bibinfo {author} {\bibfnamefont {S.}~\bibnamefont
  {Mukhopadhyay}}, \bibinfo {author} {\bibfnamefont {H.}~\bibnamefont
  {Eisaki}}, \bibinfo {author} {\bibfnamefont {S.}~\bibnamefont {Uchida}},
  \bibinfo {author} {\bibfnamefont {M.~J.}\ \bibnamefont {Lawler}}, \bibinfo
  {author} {\bibfnamefont {E.-A.}\ \bibnamefont {Kim}}, \ and\ \bibinfo
  {author} {\bibfnamefont {J.~C.}\ \bibnamefont {Davis}},\
  }\href{\doibase/10.1126/science.1248783} {\bibfield  {journal} {\bibinfo
  {journal} {Science}\ }\textbf {\bibinfo {volume} {344}},\ \bibinfo {pages}
  {612} (\bibinfo {year} {2014})}\BibitemShut {NoStop}%
\bibitem [{\citenamefont {Comin}\ \emph {\textit{et~al.}}(2014)\citenamefont
  {Comin}, \citenamefont {Frano}, \citenamefont {Yee}, \citenamefont {Yoshida},
  \citenamefont {Eisaki}, \citenamefont {Schierle}, \citenamefont {Weschke},
  \citenamefont {Sutarto}, \citenamefont {He}, \citenamefont {Soumyanarayanan},
  \citenamefont {He}, \citenamefont {Le~Tacon}, \citenamefont {Elfimov},
  \citenamefont {Hoffman}, \citenamefont {Sawatzky}, \citenamefont {Keimer},\
  and\ \citenamefont {Damascelli}}]{Comin2014_Bi2201}%
  \BibitemOpen
  \bibfield  {author} {\bibinfo {author} {\bibfnamefont {R.}~\bibnamefont
  {Comin}}, \bibinfo {author} {\bibfnamefont {A.}~\bibnamefont {Frano}},
  \bibinfo {author} {\bibfnamefont {M.~M.}\ \bibnamefont {Yee}}, \bibinfo
  {author} {\bibfnamefont {Y.}~\bibnamefont {Yoshida}}, \bibinfo {author}
  {\bibfnamefont {H.}~\bibnamefont {Eisaki}}, \bibinfo {author} {\bibfnamefont
  {E.}~\bibnamefont {Schierle}}, \bibinfo {author} {\bibfnamefont
  {E.}~\bibnamefont {Weschke}}, \bibinfo {author} {\bibfnamefont
  {R.}~\bibnamefont {Sutarto}}, \bibinfo {author} {\bibfnamefont
  {F.}~\bibnamefont {He}}, \bibinfo {author} {\bibfnamefont {A.}~\bibnamefont
  {Soumyanarayanan}}, \bibinfo {author} {\bibfnamefont {Y.}~\bibnamefont {He}},
  \bibinfo {author} {\bibfnamefont {M.}~\bibnamefont {Le~Tacon}}, \bibinfo
  {author} {\bibfnamefont {I.~S.}\ \bibnamefont {Elfimov}}, \bibinfo {author}
  {\bibfnamefont {J.~E.}\ \bibnamefont {Hoffman}}, \bibinfo {author}
  {\bibfnamefont {G.~A.}\ \bibnamefont {Sawatzky}}, \bibinfo {author}
  {\bibfnamefont {B.}~\bibnamefont {Keimer}}, \ and\ \bibinfo {author}
  {\bibfnamefont {A.}~\bibnamefont {Damascelli}},\
  }\href{\doibase/10.1126/science.1242996} {\bibfield  {journal} {\bibinfo
  {journal} {Science}\ }\textbf {\bibinfo {volume} {343}},\ \bibinfo {pages}
  {390} (\bibinfo {year} {2014})}\BibitemShut {NoStop}%
\bibitem [{\citenamefont {Caprara}\ \emph {\textit{et~al.}}(2017)\citenamefont
  {Caprara}, \citenamefont {Di~Castro}, \citenamefont {Seibold},\ and\
  \citenamefont {Grilli}}]{Caprara2017}%
  \BibitemOpen
  \bibfield  {author} {\bibinfo {author} {\bibfnamefont {S.}~\bibnamefont
  {Caprara}}, \bibinfo {author} {\bibfnamefont {C.}~\bibnamefont {Di~Castro}},
  \bibinfo {author} {\bibfnamefont {G.}~\bibnamefont {Seibold}}, \ and\
  \bibinfo {author} {\bibfnamefont {M.}~\bibnamefont {Grilli}},\
  }\href{\doibase/10.1103/PhysRevB.95.224511} {\bibfield  {journal} {\bibinfo
  {journal} {Phys. Rev. B}\ }\textbf {\bibinfo {volume} {95}},\ \bibinfo
  {pages} {224511} (\bibinfo {year} {2017})}\BibitemShut {NoStop}%
\bibitem [{\citenamefont {Blanco-Canosa}\ \emph
  {\textit{et~al.}}(2014)\citenamefont {Blanco-Canosa}, \citenamefont {Frano},
  \citenamefont {Schierle}, \citenamefont {Porras}, \citenamefont {Loew},
  \citenamefont {Minola}, \citenamefont {Bluschke}, \citenamefont {Weschke},
  \citenamefont {Keimer},\ and\ \citenamefont {Le~Tacon}}]{Santi_PRB_2014}%
  \BibitemOpen
  \bibfield  {author} {\bibinfo {author} {\bibfnamefont {S.}~\bibnamefont
  {Blanco-Canosa}}, \bibinfo {author} {\bibfnamefont {A.}~\bibnamefont
  {Frano}}, \bibinfo {author} {\bibfnamefont {E.}~\bibnamefont {Schierle}},
  \bibinfo {author} {\bibfnamefont {J.}~\bibnamefont {Porras}}, \bibinfo
  {author} {\bibfnamefont {T.}~\bibnamefont {Loew}}, \bibinfo {author}
  {\bibfnamefont {M.}~\bibnamefont {Minola}}, \bibinfo {author} {\bibfnamefont
  {M.}~\bibnamefont {Bluschke}}, \bibinfo {author} {\bibfnamefont
  {E.}~\bibnamefont {Weschke}}, \bibinfo {author} {\bibfnamefont
  {B.}~\bibnamefont {Keimer}}, \ and\ \bibinfo {author} {\bibfnamefont
  {M.}~\bibnamefont {Le~Tacon}},\ }\href{\doibase/10.1103/PhysRevB.90.054513}
  {\bibfield  {journal} {\bibinfo  {journal} {Phys. Rev. B}\ }\textbf {\bibinfo
  {volume} {90}},\ \bibinfo {pages} {054513} (\bibinfo {year}
  {2014})}\BibitemShut {NoStop}%
\bibitem [{\citenamefont {Badoux}\ \emph {\textit{et~al.}}(2016)\citenamefont
  {Badoux}, \citenamefont {Tabis}, \citenamefont {Lalibert{\'e}}, \citenamefont
  {Grissonnanche}, \citenamefont {Vignolle}, \citenamefont {Vignolles},
  \citenamefont {B{\'e}ard}, \citenamefont {Bonn}, \citenamefont {Hardy},
  \citenamefont {Liang}, \citenamefont {Doiron-Leyraud}, \citenamefont
  {Taillefer},\ and\ \citenamefont {Proust}}]{Badoux2016}%
  \BibitemOpen
  \bibfield  {author} {\bibinfo {author} {\bibfnamefont {S.}~\bibnamefont
  {Badoux}}, \bibinfo {author} {\bibfnamefont {W.}~\bibnamefont {Tabis}},
  \bibinfo {author} {\bibfnamefont {F.}~\bibnamefont {Lalibert{\'e}}}, \bibinfo
  {author} {\bibfnamefont {G.}~\bibnamefont {Grissonnanche}}, \bibinfo {author}
  {\bibfnamefont {B.}~\bibnamefont {Vignolle}}, \bibinfo {author}
  {\bibfnamefont {D.}~\bibnamefont {Vignolles}}, \bibinfo {author}
  {\bibfnamefont {J.}~\bibnamefont {B{\'e}ard}}, \bibinfo {author}
  {\bibfnamefont {D.~A.}\ \bibnamefont {Bonn}}, \bibinfo {author}
  {\bibfnamefont {W.~N.}\ \bibnamefont {Hardy}}, \bibinfo {author}
  {\bibfnamefont {R.}~\bibnamefont {Liang}}, \bibinfo {author} {\bibfnamefont
  {N.}~\bibnamefont {Doiron-Leyraud}}, \bibinfo {author} {\bibfnamefont
  {L.}~\bibnamefont {Taillefer}}, \ and\ \bibinfo {author} {\bibfnamefont
  {C.}~\bibnamefont {Proust}},\ }\href{https://doi.org/10.1038/nature16983}
  {\bibfield  {journal} {\bibinfo  {journal} {Nature}\ }\textbf {\bibinfo
  {volume} {531}},\ \bibinfo {pages} {210} (\bibinfo {year}
  {2016})}\BibitemShut {NoStop}%
\bibitem [{\citenamefont {Peng}\ \emph {\textit{et~al.}}(2018)\citenamefont
  {Peng}, \citenamefont {Fumagalli}, \citenamefont {Ding}, \citenamefont
  {Minola}, \citenamefont {Caprara}, \citenamefont {Betto}, \citenamefont
  {Bluschke}, \citenamefont {De~Luca}, \citenamefont {Kummer}, \citenamefont
  {Lefran{\c{c}}ois}, \citenamefont {Salluzzo}, \citenamefont {Suzuki},
  \citenamefont {Le~Tacon}, \citenamefont {Zhou}, \citenamefont {Brookes},
  \citenamefont {Keimer}, \citenamefont {Braicovich}, \citenamefont {Grilli},\
  and\ \citenamefont {Ghiringhelli}}]{Peng2018}%
  \BibitemOpen
  \bibfield  {author} {\bibinfo {author} {\bibfnamefont {Y.~Y.}\ \bibnamefont
  {Peng}}, \bibinfo {author} {\bibfnamefont {R.}~\bibnamefont {Fumagalli}},
  \bibinfo {author} {\bibfnamefont {Y.}~\bibnamefont {Ding}}, \bibinfo {author}
  {\bibfnamefont {M.}~\bibnamefont {Minola}}, \bibinfo {author} {\bibfnamefont
  {S.}~\bibnamefont {Caprara}}, \bibinfo {author} {\bibfnamefont
  {D.}~\bibnamefont {Betto}}, \bibinfo {author} {\bibfnamefont
  {M.}~\bibnamefont {Bluschke}}, \bibinfo {author} {\bibfnamefont {G.~M.}\
  \bibnamefont {De~Luca}}, \bibinfo {author} {\bibfnamefont {K.}~\bibnamefont
  {Kummer}}, \bibinfo {author} {\bibfnamefont {E.}~\bibnamefont
  {Lefran{\c{c}}ois}}, \bibinfo {author} {\bibfnamefont {M.}~\bibnamefont
  {Salluzzo}}, \bibinfo {author} {\bibfnamefont {H.}~\bibnamefont {Suzuki}},
  \bibinfo {author} {\bibfnamefont {M.}~\bibnamefont {Le~Tacon}}, \bibinfo
  {author} {\bibfnamefont {X.~J.}\ \bibnamefont {Zhou}}, \bibinfo {author}
  {\bibfnamefont {N.~B.}\ \bibnamefont {Brookes}}, \bibinfo {author}
  {\bibfnamefont {B.}~\bibnamefont {Keimer}}, \bibinfo {author} {\bibfnamefont
  {L.}~\bibnamefont {Braicovich}}, \bibinfo {author} {\bibfnamefont
  {M.}~\bibnamefont {Grilli}}, \ and\ \bibinfo {author} {\bibfnamefont
  {G.}~\bibnamefont {Ghiringhelli}},\
  }\href{\doibase/10.1038/s41563-018-0108-3} {\bibfield  {journal} {\bibinfo
  {journal} {Nat. Mater.}\ }\textbf {\bibinfo {volume} {17}},\ \bibinfo {pages}
  {697} (\bibinfo {year} {2018})}\BibitemShut {NoStop}%
\bibitem [{\citenamefont {Ghiringhelli}\ \emph
  {\textit{et~al.}}(2012)\citenamefont {Ghiringhelli}, \citenamefont
  {Le~Tacon}, \citenamefont {Minola}, \citenamefont {Blanco-Canosa},
  \citenamefont {Mazzoli}, \citenamefont {Brookes}, \citenamefont {De~Luca},
  \citenamefont {Frano}, \citenamefont {Hawthorn}, \citenamefont {He},
  \citenamefont {Loew}, \citenamefont {Sala}, \citenamefont {Peets},
  \citenamefont {Salluzzo}, \citenamefont {Schierle}, \citenamefont {Sutarto},
  \citenamefont {Sawatzky}, \citenamefont {Weschke}, \citenamefont {Keimer},\
  and\ \citenamefont {Braicovich}}]{Ghiringhelli2012}%
  \BibitemOpen
  \bibfield  {author} {\bibinfo {author} {\bibfnamefont {G.}~\bibnamefont
  {Ghiringhelli}}, \bibinfo {author} {\bibfnamefont {M.}~\bibnamefont
  {Le~Tacon}}, \bibinfo {author} {\bibfnamefont {M.}~\bibnamefont {Minola}},
  \bibinfo {author} {\bibfnamefont {S.}~\bibnamefont {Blanco-Canosa}}, \bibinfo
  {author} {\bibfnamefont {C.}~\bibnamefont {Mazzoli}}, \bibinfo {author}
  {\bibfnamefont {N.~B.}\ \bibnamefont {Brookes}}, \bibinfo {author}
  {\bibfnamefont {G.~M.}\ \bibnamefont {De~Luca}}, \bibinfo {author}
  {\bibfnamefont {A.}~\bibnamefont {Frano}}, \bibinfo {author} {\bibfnamefont
  {D.~G.}\ \bibnamefont {Hawthorn}}, \bibinfo {author} {\bibfnamefont
  {F.}~\bibnamefont {He}}, \bibinfo {author} {\bibfnamefont {T.}~\bibnamefont
  {Loew}}, \bibinfo {author} {\bibfnamefont {M.~M.}\ \bibnamefont {Sala}},
  \bibinfo {author} {\bibfnamefont {D.~C.}\ \bibnamefont {Peets}}, \bibinfo
  {author} {\bibfnamefont {M.}~\bibnamefont {Salluzzo}}, \bibinfo {author}
  {\bibfnamefont {E.}~\bibnamefont {Schierle}}, \bibinfo {author}
  {\bibfnamefont {R.}~\bibnamefont {Sutarto}}, \bibinfo {author} {\bibfnamefont
  {G.~A.}\ \bibnamefont {Sawatzky}}, \bibinfo {author} {\bibfnamefont
  {E.}~\bibnamefont {Weschke}}, \bibinfo {author} {\bibfnamefont
  {B.}~\bibnamefont {Keimer}}, \ and\ \bibinfo {author} {\bibfnamefont
  {L.}~\bibnamefont {Braicovich}},\ }\href{\doibase/10.1126/science.1223532}
  {\bibfield  {journal} {\bibinfo  {journal} {Science}\ }\textbf {\bibinfo
  {volume} {337}},\ \bibinfo {pages} {821} (\bibinfo {year}
  {2012})}\BibitemShut {NoStop}%
\bibitem [{\citenamefont {Chang}\ \emph {\textit{et~al.}}(2012)\citenamefont
  {Chang}, \citenamefont {Blackburn}, \citenamefont {Holmes}, \citenamefont
  {Christensen}, \citenamefont {Larsen}, \citenamefont {Mesot}, \citenamefont
  {Liang}, \citenamefont {Bonn}, \citenamefont {Hardy}, \citenamefont
  {Watenphul}, \citenamefont {Zimmermann}, \citenamefont {Forgan},\ and\
  \citenamefont {Hayden}}]{Chang2012}%
  \BibitemOpen
  \bibfield  {author} {\bibinfo {author} {\bibfnamefont {J.}~\bibnamefont
  {Chang}}, \bibinfo {author} {\bibfnamefont {E.}~\bibnamefont {Blackburn}},
  \bibinfo {author} {\bibfnamefont {A.~T.}\ \bibnamefont {Holmes}}, \bibinfo
  {author} {\bibfnamefont {N.~B.}\ \bibnamefont {Christensen}}, \bibinfo
  {author} {\bibfnamefont {J.}~\bibnamefont {Larsen}}, \bibinfo {author}
  {\bibfnamefont {J.}~\bibnamefont {Mesot}}, \bibinfo {author} {\bibfnamefont
  {R.}~\bibnamefont {Liang}}, \bibinfo {author} {\bibfnamefont {D.~A.}\
  \bibnamefont {Bonn}}, \bibinfo {author} {\bibfnamefont {W.~N.}\ \bibnamefont
  {Hardy}}, \bibinfo {author} {\bibfnamefont {A.}~\bibnamefont {Watenphul}},
  \bibinfo {author} {\bibfnamefont {M.~v.}\ \bibnamefont {Zimmermann}},
  \bibinfo {author} {\bibfnamefont {E.~M.}\ \bibnamefont {Forgan}}, \ and\
  \bibinfo {author} {\bibfnamefont {S.~M.}\ \bibnamefont {Hayden}},\
  }\href{http://dx.doi.org/10.1038/nphys2456} {\bibfield  {journal} {\bibinfo
  {journal} {Nat. Phys.}\ }\textbf {\bibinfo {volume} {8}},\ \bibinfo {pages}
  {871} (\bibinfo {year} {2012})}\BibitemShut {NoStop}%
\bibitem [{\citenamefont {H\"ucker}\ \emph {\textit{et~al.}}(2014)\citenamefont
  {H\"ucker}, \citenamefont {Christensen}, \citenamefont {Holmes},
  \citenamefont {Blackburn}, \citenamefont {Forgan}, \citenamefont {Liang},
  \citenamefont {Bonn}, \citenamefont {Hardy}, \citenamefont {Gutowski},
  \citenamefont {Zimmermann}, \citenamefont {Hayden},\ and\ \citenamefont
  {Chang}}]{Huecker2014}%
  \BibitemOpen
  \bibfield  {author} {\bibinfo {author} {\bibfnamefont {M.}~\bibnamefont
  {H\"ucker}}, \bibinfo {author} {\bibfnamefont {N.~B.}\ \bibnamefont
  {Christensen}}, \bibinfo {author} {\bibfnamefont {A.~T.}\ \bibnamefont
  {Holmes}}, \bibinfo {author} {\bibfnamefont {E.}~\bibnamefont {Blackburn}},
  \bibinfo {author} {\bibfnamefont {E.~M.}\ \bibnamefont {Forgan}}, \bibinfo
  {author} {\bibfnamefont {R.}~\bibnamefont {Liang}}, \bibinfo {author}
  {\bibfnamefont {D.~A.}\ \bibnamefont {Bonn}}, \bibinfo {author}
  {\bibfnamefont {W.~N.}\ \bibnamefont {Hardy}}, \bibinfo {author}
  {\bibfnamefont {O.}~\bibnamefont {Gutowski}}, \bibinfo {author}
  {\bibfnamefont {M.~v.}\ \bibnamefont {Zimmermann}}, \bibinfo {author}
  {\bibfnamefont {S.~M.}\ \bibnamefont {Hayden}}, \ and\ \bibinfo {author}
  {\bibfnamefont {J.}~\bibnamefont {Chang}},\
  }\href{\doibase/10.1103/PhysRevB.90.054514} {\bibfield  {journal} {\bibinfo
  {journal} {Phys. Rev. B}\ }\textbf {\bibinfo {volume} {90}},\ \bibinfo
  {pages} {054514} (\bibinfo {year} {2014})}\BibitemShut {NoStop}%
\bibitem [{\citenamefont {da~Silva~Neto}\ \emph
  {\textit{et~al.}}(2014)\citenamefont {da~Silva~Neto}, \citenamefont
  {Aynajian}, \citenamefont {Frano}, \citenamefont {Comin}, \citenamefont
  {Schierle}, \citenamefont {Weschke}, \citenamefont {Gyenis}, \citenamefont
  {Wen}, \citenamefont {Schneeloch}, \citenamefont {Xu}, \citenamefont {Ono},
  \citenamefont {Gu}, \citenamefont {Le~Tacon},\ and\ \citenamefont
  {Yazdani}}]{Neto_2014_Bi2212}%
  \BibitemOpen
  \bibfield  {author} {\bibinfo {author} {\bibfnamefont {E.~H.}\ \bibnamefont
  {da~Silva~Neto}}, \bibinfo {author} {\bibfnamefont {P.}~\bibnamefont
  {Aynajian}}, \bibinfo {author} {\bibfnamefont {A.}~\bibnamefont {Frano}},
  \bibinfo {author} {\bibfnamefont {R.}~\bibnamefont {Comin}}, \bibinfo
  {author} {\bibfnamefont {E.}~\bibnamefont {Schierle}}, \bibinfo {author}
  {\bibfnamefont {E.}~\bibnamefont {Weschke}}, \bibinfo {author} {\bibfnamefont
  {A.}~\bibnamefont {Gyenis}}, \bibinfo {author} {\bibfnamefont
  {J.}~\bibnamefont {Wen}}, \bibinfo {author} {\bibfnamefont {J.}~\bibnamefont
  {Schneeloch}}, \bibinfo {author} {\bibfnamefont {Z.}~\bibnamefont {Xu}},
  \bibinfo {author} {\bibfnamefont {S.}~\bibnamefont {Ono}}, \bibinfo {author}
  {\bibfnamefont {G.}~\bibnamefont {Gu}}, \bibinfo {author} {\bibfnamefont
  {M.}~\bibnamefont {Le~Tacon}}, \ and\ \bibinfo {author} {\bibfnamefont
  {A.}~\bibnamefont {Yazdani}},\ }\href{\doibase/10.1126/science.1243479}
  {\bibfield  {journal} {\bibinfo  {journal} {Science}\ }\textbf {\bibinfo
  {volume} {343}},\ \bibinfo {pages} {393} (\bibinfo {year}
  {2014})}\BibitemShut {NoStop}%
\bibitem [{\citenamefont {Tabis}\ \emph {\textit{et~al.}}(2014)\citenamefont
  {Tabis}, \citenamefont {Li}, \citenamefont {Tacon}, \citenamefont
  {Braicovich}, \citenamefont {Kreyssig}, \citenamefont {Minola}, \citenamefont
  {Dellea}, \citenamefont {Weschke}, \citenamefont {Veit}, \citenamefont
  {Ramazanoglu}, \citenamefont {Goldman}, \citenamefont {Schmitt},
  \citenamefont {Ghiringhelli}, \citenamefont {Barisic}, \citenamefont {Chan},
  \citenamefont {Dorow}, \citenamefont {Yu}, \citenamefont {Zhao},
  \citenamefont {Keimer},\ and\ \citenamefont {Greven}}]{Tabis_2014_Hg1201}%
  \BibitemOpen
  \bibfield  {author} {\bibinfo {author} {\bibfnamefont {W.}~\bibnamefont
  {Tabis}}, \bibinfo {author} {\bibfnamefont {Y.}~\bibnamefont {Li}}, \bibinfo
  {author} {\bibfnamefont {M.~L.}\ \bibnamefont {Tacon}}, \bibinfo {author}
  {\bibfnamefont {L.}~\bibnamefont {Braicovich}}, \bibinfo {author}
  {\bibfnamefont {A.}~\bibnamefont {Kreyssig}}, \bibinfo {author}
  {\bibfnamefont {M.}~\bibnamefont {Minola}}, \bibinfo {author} {\bibfnamefont
  {G.}~\bibnamefont {Dellea}}, \bibinfo {author} {\bibfnamefont
  {E.}~\bibnamefont {Weschke}}, \bibinfo {author} {\bibfnamefont {M.~J.}\
  \bibnamefont {Veit}}, \bibinfo {author} {\bibfnamefont {M.}~\bibnamefont
  {Ramazanoglu}}, \bibinfo {author} {\bibfnamefont {A.~I.}\ \bibnamefont
  {Goldman}}, \bibinfo {author} {\bibfnamefont {T.}~\bibnamefont {Schmitt}},
  \bibinfo {author} {\bibfnamefont {G.}~\bibnamefont {Ghiringhelli}}, \bibinfo
  {author} {\bibfnamefont {N.}~\bibnamefont {Barisic}}, \bibinfo {author}
  {\bibfnamefont {M.~K.}\ \bibnamefont {Chan}}, \bibinfo {author}
  {\bibfnamefont {C.~J.}\ \bibnamefont {Dorow}}, \bibinfo {author}
  {\bibfnamefont {G.}~\bibnamefont {Yu}}, \bibinfo {author} {\bibfnamefont
  {X.}~\bibnamefont {Zhao}}, \bibinfo {author} {\bibfnamefont {B.}~\bibnamefont
  {Keimer}}, \ and\ \bibinfo {author} {\bibfnamefont {M.}~\bibnamefont
  {Greven}},\ }\href{https://doi.org/10.1038/ncomms6875} {\bibfield  {journal}
  {\bibinfo  {journal} {Nat. Commun.}\ }\textbf {\bibinfo {volume} {5}},\
  \bibinfo {pages} {5875} (\bibinfo {year} {2014})}\BibitemShut {NoStop}%
\bibitem [{\citenamefont {Tabis}\ \emph {\textit{et~al.}}(2017)\citenamefont
  {Tabis}, \citenamefont {Yu}, \citenamefont {Bialo}, \citenamefont {Bluschke},
  \citenamefont {Kolodziej}, \citenamefont {Kozlowski}, \citenamefont
  {Blackburn}, \citenamefont {Sen}, \citenamefont {Forgan}, \citenamefont
  {Zimmermann}, \citenamefont {Tang}, \citenamefont {Weschke}, \citenamefont
  {Vignolle}, \citenamefont {Hepting}, \citenamefont {Gretarsson},
  \citenamefont {Sutarto}, \citenamefont {He}, \citenamefont {Le~Tacon},
  \citenamefont {Bari\ifmmode \check{s}\else \v{s}\fi{}i\ifmmode~\acute{c}\else
  \'{c}\fi{}}, \citenamefont {Yu},\ and\ \citenamefont {Greven}}]{Tabis2017}%
  \BibitemOpen
  \bibfield  {author} {\bibinfo {author} {\bibfnamefont {W.}~\bibnamefont
  {Tabis}}, \bibinfo {author} {\bibfnamefont {B.}~\bibnamefont {Yu}}, \bibinfo
  {author} {\bibfnamefont {I.}~\bibnamefont {Bialo}}, \bibinfo {author}
  {\bibfnamefont {M.}~\bibnamefont {Bluschke}}, \bibinfo {author}
  {\bibfnamefont {T.}~\bibnamefont {Kolodziej}}, \bibinfo {author}
  {\bibfnamefont {A.}~\bibnamefont {Kozlowski}}, \bibinfo {author}
  {\bibfnamefont {E.}~\bibnamefont {Blackburn}}, \bibinfo {author}
  {\bibfnamefont {K.}~\bibnamefont {Sen}}, \bibinfo {author} {\bibfnamefont
  {E.~M.}\ \bibnamefont {Forgan}}, \bibinfo {author} {\bibfnamefont {M.~v.}\
  \bibnamefont {Zimmermann}}, \bibinfo {author} {\bibfnamefont
  {Y.}~\bibnamefont {Tang}}, \bibinfo {author} {\bibfnamefont {E.}~\bibnamefont
  {Weschke}}, \bibinfo {author} {\bibfnamefont {B.}~\bibnamefont {Vignolle}},
  \bibinfo {author} {\bibfnamefont {M.}~\bibnamefont {Hepting}}, \bibinfo
  {author} {\bibfnamefont {H.}~\bibnamefont {Gretarsson}}, \bibinfo {author}
  {\bibfnamefont {R.}~\bibnamefont {Sutarto}}, \bibinfo {author} {\bibfnamefont
  {F.}~\bibnamefont {He}}, \bibinfo {author} {\bibfnamefont {M.}~\bibnamefont
  {Le~Tacon}}, \bibinfo {author} {\bibfnamefont {N.}~\bibnamefont {Bari\ifmmode
  \check{s}\else \v{s}\fi{}i\ifmmode~\acute{c}\else \'{c}\fi{}}}, \bibinfo
  {author} {\bibfnamefont {G.}~\bibnamefont {Yu}}, \ and\ \bibinfo {author}
  {\bibfnamefont {M.}~\bibnamefont {Greven}},\
  }\href{\doibase/10.1103/PhysRevB.96.134510} {\bibfield  {journal} {\bibinfo
  {journal} {Phys. Rev. B}\ }\textbf {\bibinfo {volume} {96}},\ \bibinfo
  {pages} {134510} (\bibinfo {year} {2017})}\BibitemShut {NoStop}%
\bibitem [{\citenamefont {Wang}\ and\ \citenamefont
  {Chubukov}(2014)}]{Wang2014}%
  \BibitemOpen
  \bibfield  {author} {\bibinfo {author} {\bibfnamefont {Y.}~\bibnamefont
  {Wang}}\ and\ \bibinfo {author} {\bibfnamefont {A.}~\bibnamefont
  {Chubukov}},\ }\href{\doibase/10.1103/PhysRevB.90.035149} {\bibfield
  {journal} {\bibinfo  {journal} {Phys. Rev. B}\ }\textbf {\bibinfo {volume}
  {90}},\ \bibinfo {pages} {035149} (\bibinfo {year} {2014})}\BibitemShut
  {NoStop}%
\bibitem [{\citenamefont {Efetov}\ \emph {\textit{et~al.}}(2013)\citenamefont
  {Efetov}, \citenamefont {Meier},\ and\ \citenamefont
  {P{\'e}pin}}]{Efetov2013}%
  \BibitemOpen
  \bibfield  {author} {\bibinfo {author} {\bibfnamefont {K.~B.}\ \bibnamefont
  {Efetov}}, \bibinfo {author} {\bibfnamefont {H.}~\bibnamefont {Meier}}, \
  and\ \bibinfo {author} {\bibfnamefont {C.}~\bibnamefont {P{\'e}pin}},\
  }\href{https://doi.org/10.1038/nphys2641} {\bibfield  {journal} {\bibinfo
  {journal} {Nat. Phys.}\ }\textbf {\bibinfo {volume} {9}},\ \bibinfo {pages}
  {442} (\bibinfo {year} {2013})}\BibitemShut {NoStop}%
\bibitem [{\citenamefont {Montiel}\ \emph {\textit{et~al.}}(2017)\citenamefont
  {Montiel}, \citenamefont {Kloss},\ and\ \citenamefont
  {P\'epin}}]{Montiel2017}%
  \BibitemOpen
  \bibfield  {author} {\bibinfo {author} {\bibfnamefont {X.}~\bibnamefont
  {Montiel}}, \bibinfo {author} {\bibfnamefont {T.}~\bibnamefont {Kloss}}, \
  and\ \bibinfo {author} {\bibfnamefont {C.}~\bibnamefont {P\'epin}},\
  }\href{\doibase/10.1103/PhysRevB.95.104510} {\bibfield  {journal} {\bibinfo
  {journal} {Phys. Rev. B}\ }\textbf {\bibinfo {volume} {95}},\ \bibinfo
  {pages} {104510} (\bibinfo {year} {2017})}\BibitemShut {NoStop}%
\bibitem [{\citenamefont {Kanigel}\ \emph {\textit{et~al.}}(2002)\citenamefont
  {Kanigel}, \citenamefont {Keren}, \citenamefont {Eckstein}, \citenamefont
  {Knizhnik}, \citenamefont {Lord},\ and\ \citenamefont
  {Amato}}]{Kanigel2002_PRL}%
  \BibitemOpen
  \bibfield  {author} {\bibinfo {author} {\bibfnamefont {A.}~\bibnamefont
  {Kanigel}}, \bibinfo {author} {\bibfnamefont {A.}~\bibnamefont {Keren}},
  \bibinfo {author} {\bibfnamefont {Y.}~\bibnamefont {Eckstein}}, \bibinfo
  {author} {\bibfnamefont {A.}~\bibnamefont {Knizhnik}}, \bibinfo {author}
  {\bibfnamefont {J.~S.}\ \bibnamefont {Lord}}, \ and\ \bibinfo {author}
  {\bibfnamefont {A.}~\bibnamefont {Amato}},\
  }\href{\doibase/10.1103/PhysRevLett.88.137003} {\bibfield  {journal}
  {\bibinfo  {journal} {Phys. Rev. Lett.}\ }\textbf {\bibinfo {volume} {88}},\
  \bibinfo {pages} {137003} (\bibinfo {year} {2002})}\BibitemShut {NoStop}%
\bibitem [{\citenamefont {Lubashevsky}\ and\ \citenamefont
  {Keren}(2008)}]{Lubashevsky_PRB_2008}%
  \BibitemOpen
  \bibfield  {author} {\bibinfo {author} {\bibfnamefont {Y.}~\bibnamefont
  {Lubashevsky}}\ and\ \bibinfo {author} {\bibfnamefont {A.}~\bibnamefont
  {Keren}},\ }\href{\doibase/10.1103/PhysRevB.78.020505} {\bibfield  {journal}
  {\bibinfo  {journal} {Phys. Rev. B}\ }\textbf {\bibinfo {volume} {78}},\
  \bibinfo {pages} {020505(R)} (\bibinfo {year} {2008})}\BibitemShut {NoStop}%
\bibitem [{\citenamefont {Cvitani\ifmmode~\acute{c}\else \'{c}\fi{}}\ \emph
  {\textit{et~al.}}(2014)\citenamefont {Cvitani\ifmmode~\acute{c}\else
  \'{c}\fi{}}, \citenamefont {Pelc}, \citenamefont {Po\ifmmode~\check{z}\else
  \v{z}\fi{}ek}, \citenamefont {Amit},\ and\ \citenamefont
  {Keren}}]{Cvitanic2014}%
  \BibitemOpen
  \bibfield  {author} {\bibinfo {author} {\bibfnamefont {T.}~\bibnamefont
  {Cvitani\ifmmode~\acute{c}\else \'{c}\fi{}}}, \bibinfo {author}
  {\bibfnamefont {D.}~\bibnamefont {Pelc}}, \bibinfo {author} {\bibfnamefont
  {M.}~\bibnamefont {Po\ifmmode~\check{z}\else \v{z}\fi{}ek}}, \bibinfo
  {author} {\bibfnamefont {E.}~\bibnamefont {Amit}}, \ and\ \bibinfo {author}
  {\bibfnamefont {A.}~\bibnamefont {Keren}},\
  }\href{\doibase/10.1103/PhysRevB.90.054508} {\bibfield  {journal} {\bibinfo
  {journal} {Phys. Rev. B}\ }\textbf {\bibinfo {volume} {90}},\ \bibinfo
  {pages} {054508} (\bibinfo {year} {2014})}\BibitemShut {NoStop}%
\bibitem [{\citenamefont {Ofer}\ \emph {\textit{et~al.}}(2006)\citenamefont
  {Ofer}, \citenamefont {Bazalitsky}, \citenamefont {Kanigel}, \citenamefont
  {Keren}, \citenamefont {Auerbach}, \citenamefont {Lord},\ and\ \citenamefont
  {Amato}}]{Ofer2006_magnetic_analog}%
  \BibitemOpen
  \bibfield  {author} {\bibinfo {author} {\bibfnamefont {R.}~\bibnamefont
  {Ofer}}, \bibinfo {author} {\bibfnamefont {G.}~\bibnamefont {Bazalitsky}},
  \bibinfo {author} {\bibfnamefont {A.}~\bibnamefont {Kanigel}}, \bibinfo
  {author} {\bibfnamefont {A.}~\bibnamefont {Keren}}, \bibinfo {author}
  {\bibfnamefont {A.}~\bibnamefont {Auerbach}}, \bibinfo {author}
  {\bibfnamefont {J.~S.}\ \bibnamefont {Lord}}, \ and\ \bibinfo {author}
  {\bibfnamefont {A.}~\bibnamefont {Amato}},\
  }\href{\doibase/10.1103/PhysRevB.74.220508} {\bibfield  {journal} {\bibinfo
  {journal} {Phys. Rev. B}\ }\textbf {\bibinfo {volume} {74}},\ \bibinfo
  {pages} {220508(R)} (\bibinfo {year} {2006})}\BibitemShut {NoStop}%
\bibitem [{\citenamefont {Yamada}\ \emph {\textit{et~al.}}(1998)\citenamefont
  {Yamada}, \citenamefont {Lee}, \citenamefont {Kurahashi}, \citenamefont
  {Wada}, \citenamefont {Wakimoto}, \citenamefont {Ueki}, \citenamefont
  {Kimura}, \citenamefont {Endoh}, \citenamefont {Hosoya}, \citenamefont
  {Shirane}, \citenamefont {Birgeneau}, \citenamefont {Greven}, \citenamefont
  {Kastner},\ and\ \citenamefont {Kim}}]{Yamada_PRB_1998}%
  \BibitemOpen
  \bibfield  {author} {\bibinfo {author} {\bibfnamefont {K.}~\bibnamefont
  {Yamada}}, \bibinfo {author} {\bibfnamefont {C.~H.}\ \bibnamefont {Lee}},
  \bibinfo {author} {\bibfnamefont {K.}~\bibnamefont {Kurahashi}}, \bibinfo
  {author} {\bibfnamefont {J.}~\bibnamefont {Wada}}, \bibinfo {author}
  {\bibfnamefont {S.}~\bibnamefont {Wakimoto}}, \bibinfo {author}
  {\bibfnamefont {S.}~\bibnamefont {Ueki}}, \bibinfo {author} {\bibfnamefont
  {H.}~\bibnamefont {Kimura}}, \bibinfo {author} {\bibfnamefont
  {Y.}~\bibnamefont {Endoh}}, \bibinfo {author} {\bibfnamefont
  {S.}~\bibnamefont {Hosoya}}, \bibinfo {author} {\bibfnamefont
  {G.}~\bibnamefont {Shirane}}, \bibinfo {author} {\bibfnamefont {R.~J.}\
  \bibnamefont {Birgeneau}}, \bibinfo {author} {\bibfnamefont {M.}~\bibnamefont
  {Greven}}, \bibinfo {author} {\bibfnamefont {M.~A.}\ \bibnamefont {Kastner}},
  \ and\ \bibinfo {author} {\bibfnamefont {Y.~J.}\ \bibnamefont {Kim}},\
  }\href{\doibase/10.1103/PhysRevB.57.6165} {\bibfield  {journal} {\bibinfo
  {journal} {Phys. Rev. B}\ }\textbf {\bibinfo {volume} {57}},\ \bibinfo
  {pages} {6165} (\bibinfo {year} {1998})}\BibitemShut {NoStop}%
\bibitem [{\citenamefont {Achkar}\ \emph {\textit{et~al.}}(2016)\citenamefont
  {Achkar}, \citenamefont {Zwiebler}, \citenamefont {McMahon}, \citenamefont
  {He}, \citenamefont {Sutarto}, \citenamefont {Djianto}, \citenamefont {Hao},
  \citenamefont {Gingras}, \citenamefont {H{\"u}cker}, \citenamefont {Gu},
  \citenamefont {Revcolevschi}, \citenamefont {Zhang}, \citenamefont {Kim},
  \citenamefont {Geck},\ and\ \citenamefont {Hawthorn}}]{Achkar2016}%
  \BibitemOpen
  \bibfield  {author} {\bibinfo {author} {\bibfnamefont {A.~J.}\ \bibnamefont
  {Achkar}}, \bibinfo {author} {\bibfnamefont {M.}~\bibnamefont {Zwiebler}},
  \bibinfo {author} {\bibfnamefont {C.}~\bibnamefont {McMahon}}, \bibinfo
  {author} {\bibfnamefont {F.}~\bibnamefont {He}}, \bibinfo {author}
  {\bibfnamefont {R.}~\bibnamefont {Sutarto}}, \bibinfo {author} {\bibfnamefont
  {I.}~\bibnamefont {Djianto}}, \bibinfo {author} {\bibfnamefont
  {Z.}~\bibnamefont {Hao}}, \bibinfo {author} {\bibfnamefont {M.~J.~P.}\
  \bibnamefont {Gingras}}, \bibinfo {author} {\bibfnamefont {M.}~\bibnamefont
  {H{\"u}cker}}, \bibinfo {author} {\bibfnamefont {G.~D.}\ \bibnamefont {Gu}},
  \bibinfo {author} {\bibfnamefont {A.}~\bibnamefont {Revcolevschi}}, \bibinfo
  {author} {\bibfnamefont {H.}~\bibnamefont {Zhang}}, \bibinfo {author}
  {\bibfnamefont {Y.-J.}\ \bibnamefont {Kim}}, \bibinfo {author} {\bibfnamefont
  {J.}~\bibnamefont {Geck}}, \ and\ \bibinfo {author} {\bibfnamefont {D.~G.}\
  \bibnamefont {Hawthorn}},\ }\href{\doibase/10.1126/science.aad1824}
  {\bibfield  {journal} {\bibinfo  {journal} {Science}\ }\textbf {\bibinfo
  {volume} {351}},\ \bibinfo {pages} {576} (\bibinfo {year}
  {2016})}\BibitemShut {NoStop}%
\bibitem [{\citenamefont {Miao}\ \emph {\textit{et~al.}}(2017)\citenamefont
  {Miao}, \citenamefont {Lorenzana}, \citenamefont {Seibold}, \citenamefont
  {Peng}, \citenamefont {Amorese}, \citenamefont {Yakhou-Harris}, \citenamefont
  {Kummer}, \citenamefont {Brookes}, \citenamefont {Konik}, \citenamefont
  {Thampy}, \citenamefont {Gu}, \citenamefont {Ghiringhelli}, \citenamefont
  {Braicovich},\ and\ \citenamefont {Dean}}]{Miao2017}%
  \BibitemOpen
  \bibfield  {author} {\bibinfo {author} {\bibfnamefont {H.}~\bibnamefont
  {Miao}}, \bibinfo {author} {\bibfnamefont {J.}~\bibnamefont {Lorenzana}},
  \bibinfo {author} {\bibfnamefont {G.}~\bibnamefont {Seibold}}, \bibinfo
  {author} {\bibfnamefont {Y.~Y.}\ \bibnamefont {Peng}}, \bibinfo {author}
  {\bibfnamefont {A.}~\bibnamefont {Amorese}}, \bibinfo {author} {\bibfnamefont
  {F.}~\bibnamefont {Yakhou-Harris}}, \bibinfo {author} {\bibfnamefont
  {K.}~\bibnamefont {Kummer}}, \bibinfo {author} {\bibfnamefont {N.~B.}\
  \bibnamefont {Brookes}}, \bibinfo {author} {\bibfnamefont {R.~M.}\
  \bibnamefont {Konik}}, \bibinfo {author} {\bibfnamefont {V.}~\bibnamefont
  {Thampy}}, \bibinfo {author} {\bibfnamefont {G.~D.}\ \bibnamefont {Gu}},
  \bibinfo {author} {\bibfnamefont {G.}~\bibnamefont {Ghiringhelli}}, \bibinfo
  {author} {\bibfnamefont {L.}~\bibnamefont {Braicovich}}, \ and\ \bibinfo
  {author} {\bibfnamefont {M.~P.~M.}\ \bibnamefont {Dean}},\
  }\href{\doibase/10.1073/pnas.1708549114} {\bibfield  {journal} {\bibinfo
  {journal} {Proc. Natl. Acad. Sci. USA}\ }\textbf {\bibinfo {volume} {114}},\
  \bibinfo {pages} {12430} (\bibinfo {year} {2017})}\BibitemShut {NoStop}%
\bibitem [{\citenamefont {Pavarini}\ \emph {\textit{et~al.}}(2001)\citenamefont
  {Pavarini}, \citenamefont {Dasgupta}, \citenamefont {Saha-Dasgupta},
  \citenamefont {Jepsen},\ and\ \citenamefont {Andersen}}]{Pavarini2001}%
  \BibitemOpen
  \bibfield  {author} {\bibinfo {author} {\bibfnamefont {E.}~\bibnamefont
  {Pavarini}}, \bibinfo {author} {\bibfnamefont {I.}~\bibnamefont {Dasgupta}},
  \bibinfo {author} {\bibfnamefont {T.}~\bibnamefont {Saha-Dasgupta}}, \bibinfo
  {author} {\bibfnamefont {O.}~\bibnamefont {Jepsen}}, \ and\ \bibinfo {author}
  {\bibfnamefont {O.~K.}\ \bibnamefont {Andersen}},\
  }\href{\doibase/10.1103/PhysRevLett.87.047003} {\bibfield  {journal}
  {\bibinfo  {journal} {Phys. Rev. Lett.}\ }\textbf {\bibinfo {volume} {87}},\
  \bibinfo {pages} {047003} (\bibinfo {year} {2001})}\BibitemShut {NoStop}%
\bibitem [{\citenamefont {Petit}\ and\ \citenamefont
  {Lepetit}(2009)}]{Petit2009}%
  \BibitemOpen
  \bibfield  {author} {\bibinfo {author} {\bibfnamefont {S.}~\bibnamefont
  {Petit}}\ and\ \bibinfo {author} {\bibfnamefont {M.-B.}\ \bibnamefont
  {Lepetit}},\ }\href{http://stacks.iop.org/0295-5075/87/i=6/a=67005}
  {\bibfield  {journal} {\bibinfo  {journal} {Europhys. Lett.}\ }\textbf
  {\bibinfo {volume} {87}},\ \bibinfo {pages} {67005} (\bibinfo {year}
  {2009})}\BibitemShut {NoStop}%
\bibitem [{\citenamefont {Peng}\ \emph {\textit{et~al.}}(2017)\citenamefont
  {Peng}, \citenamefont {Dellea}, \citenamefont {Minola}, \citenamefont
  {Conni}, \citenamefont {Amorese}, \citenamefont {Di~Castro}, \citenamefont
  {De~Luca}, \citenamefont {Kummer}, \citenamefont {Salluzzo}, \citenamefont
  {Sun}, \citenamefont {Zhou}, \citenamefont {Balestrino}, \citenamefont
  {Le~Tacon}, \citenamefont {Keimer}, \citenamefont {Braicovich}, \citenamefont
  {Brookes},\ and\ \citenamefont {Ghiringhelli}}]{Peng2017}%
  \BibitemOpen
  \bibfield  {author} {\bibinfo {author} {\bibfnamefont {Y.~Y.}\ \bibnamefont
  {Peng}}, \bibinfo {author} {\bibfnamefont {G.}~\bibnamefont {Dellea}},
  \bibinfo {author} {\bibfnamefont {M.}~\bibnamefont {Minola}}, \bibinfo
  {author} {\bibfnamefont {M.}~\bibnamefont {Conni}}, \bibinfo {author}
  {\bibfnamefont {A.}~\bibnamefont {Amorese}}, \bibinfo {author} {\bibfnamefont
  {D.}~\bibnamefont {Di~Castro}}, \bibinfo {author} {\bibfnamefont {G.~M.}\
  \bibnamefont {De~Luca}}, \bibinfo {author} {\bibfnamefont {K.}~\bibnamefont
  {Kummer}}, \bibinfo {author} {\bibfnamefont {M.}~\bibnamefont {Salluzzo}},
  \bibinfo {author} {\bibfnamefont {X.}~\bibnamefont {Sun}}, \bibinfo {author}
  {\bibfnamefont {X.~J.}\ \bibnamefont {Zhou}}, \bibinfo {author}
  {\bibfnamefont {G.}~\bibnamefont {Balestrino}}, \bibinfo {author}
  {\bibfnamefont {M.}~\bibnamefont {Le~Tacon}}, \bibinfo {author}
  {\bibfnamefont {B.}~\bibnamefont {Keimer}}, \bibinfo {author} {\bibfnamefont
  {L.}~\bibnamefont {Braicovich}}, \bibinfo {author} {\bibfnamefont {N.~B.}\
  \bibnamefont {Brookes}}, \ and\ \bibinfo {author} {\bibfnamefont
  {G.}~\bibnamefont {Ghiringhelli}},\ }\href{https://doi.org/10.1038/nphys4248}
  {\bibfield  {journal} {\bibinfo  {journal} {Nat. Phys.}\ }\textbf {\bibinfo
  {volume} {13}},\ \bibinfo {pages} {1201} (\bibinfo {year}
  {2017})}\BibitemShut {NoStop}%
\bibitem [{\citenamefont {Bogdanov}\ \emph {\textit{et~al.}}()\citenamefont
  {Bogdanov}, \citenamefont {Manni}, \citenamefont {Sharma}, \citenamefont
  {Gunnarsson},\ and\ \citenamefont {Alavi}}]{Bogdanov2018}%
  \BibitemOpen
  \bibfield  {author} {\bibinfo {author} {\bibfnamefont {N.~A.}\ \bibnamefont
  {Bogdanov}}, \bibinfo {author} {\bibfnamefont {G.~L.}\ \bibnamefont {Manni}},
  \bibinfo {author} {\bibfnamefont {S.}~\bibnamefont {Sharma}}, \bibinfo
  {author} {\bibfnamefont {O.}~\bibnamefont {Gunnarsson}}, \ and\ \bibinfo
  {author} {\bibfnamefont {A.}~\bibnamefont {Alavi}},\
  }\href{https://arxiv.org/abs/1803.07026} {\bibinfo  {journal} {Preprint at
  https://arxiv.org/abs/1803.07026 (2018).}\ }\BibitemShut {NoStop}%
\bibitem [{\citenamefont {Goldschmidt}\ \emph
  {\textit{et~al.}}(1993)\citenamefont {Goldschmidt}, \citenamefont {Reisner},
  \citenamefont {Direktovitch}, \citenamefont {Knizhnik}, \citenamefont
  {Gartstein}, \citenamefont {Kimmel},\ and\ \citenamefont
  {Eckstein}}]{Goldschmidt1993}%
  \BibitemOpen
\bibfield  {journal} {  }\bibfield  {author} {\bibinfo {author} {\bibfnamefont
  {D.}~\bibnamefont {Goldschmidt}}, \bibinfo {author} {\bibfnamefont {G.~M.}\
  \bibnamefont {Reisner}}, \bibinfo {author} {\bibfnamefont {Y.}~\bibnamefont
  {Direktovitch}}, \bibinfo {author} {\bibfnamefont {A.}~\bibnamefont
  {Knizhnik}}, \bibinfo {author} {\bibfnamefont {E.}~\bibnamefont {Gartstein}},
  \bibinfo {author} {\bibfnamefont {G.}~\bibnamefont {Kimmel}}, \ and\ \bibinfo
  {author} {\bibfnamefont {Y.}~\bibnamefont {Eckstein}},\
  }\href{\doibase/10.1103/PhysRevB.48.532} {\bibfield  {journal} {\bibinfo
  {journal} {Phys. Rev. B}\ }\textbf {\bibinfo {volume} {48}},\ \bibinfo
  {pages} {532} (\bibinfo {year} {1993})}\BibitemShut {NoStop}%
\bibitem [{\citenamefont {Chmaissem}\ \emph
  {\textit{et~al.}}(1999)\citenamefont {Chmaissem}, \citenamefont {Jorgensen},
  \citenamefont {Short}, \citenamefont {Knizhnik}, \citenamefont {Eckstein},\
  and\ \citenamefont {Shaked}}]{Chmaissem1999}%
  \BibitemOpen
  \bibfield  {author} {\bibinfo {author} {\bibfnamefont {O.}~\bibnamefont
  {Chmaissem}}, \bibinfo {author} {\bibfnamefont {J.~D.}\ \bibnamefont
  {Jorgensen}}, \bibinfo {author} {\bibfnamefont {S.}~\bibnamefont {Short}},
  \bibinfo {author} {\bibfnamefont {A.}~\bibnamefont {Knizhnik}}, \bibinfo
  {author} {\bibfnamefont {Y.}~\bibnamefont {Eckstein}}, \ and\ \bibinfo
  {author} {\bibfnamefont {H.}~\bibnamefont {Shaked}},\
  }\href{http://dx.doi.org/10.1038/16209} {\bibfield  {journal} {\bibinfo
  {journal} {Nature}\ }\textbf {\bibinfo {volume} {397}},\ \bibinfo {pages}
  {45} (\bibinfo {year} {1999})}\BibitemShut {NoStop}%
\bibitem [{\citenamefont {Ofer}\ \emph {\textit{et~al.}}(2008)\citenamefont
  {Ofer}, \citenamefont {Keren}, \citenamefont {Chmaissem},\ and\ \citenamefont
  {Amato}}]{Ofer2008}%
  \BibitemOpen
  \bibfield  {author} {\bibinfo {author} {\bibfnamefont {R.}~\bibnamefont
  {Ofer}}, \bibinfo {author} {\bibfnamefont {A.}~\bibnamefont {Keren}},
  \bibinfo {author} {\bibfnamefont {O.}~\bibnamefont {Chmaissem}}, \ and\
  \bibinfo {author} {\bibfnamefont {A.}~\bibnamefont {Amato}},\
  }\href{\doibase/10.1103/PhysRevB.78.140508} {\bibfield  {journal} {\bibinfo
  {journal} {Phys. Rev. B}\ }\textbf {\bibinfo {volume} {78}},\ \bibinfo
  {pages} {140508(R)} (\bibinfo {year} {2008})}\BibitemShut {NoStop}%
\bibitem [{\citenamefont {Ellis}\ \emph {\textit{et~al.}}(2015)\citenamefont
  {Ellis}, \citenamefont {Huang}, \citenamefont {Olalde-Velasco}, \citenamefont
  {Dantz}, \citenamefont {Pelliciari}, \citenamefont {Drachuck}, \citenamefont
  {Ofer}, \citenamefont {Bazalitsky}, \citenamefont {Berger}, \citenamefont
  {Schmitt},\ and\ \citenamefont {Keren}}]{Ellis2015_RIXS}%
  \BibitemOpen
  \bibfield  {author} {\bibinfo {author} {\bibfnamefont {D.~S.}\ \bibnamefont
  {Ellis}}, \bibinfo {author} {\bibfnamefont {Y.-B.}\ \bibnamefont {Huang}},
  \bibinfo {author} {\bibfnamefont {P.}~\bibnamefont {Olalde-Velasco}},
  \bibinfo {author} {\bibfnamefont {M.}~\bibnamefont {Dantz}}, \bibinfo
  {author} {\bibfnamefont {J.}~\bibnamefont {Pelliciari}}, \bibinfo {author}
  {\bibfnamefont {G.}~\bibnamefont {Drachuck}}, \bibinfo {author}
  {\bibfnamefont {R.}~\bibnamefont {Ofer}}, \bibinfo {author} {\bibfnamefont
  {G.}~\bibnamefont {Bazalitsky}}, \bibinfo {author} {\bibfnamefont
  {J.}~\bibnamefont {Berger}}, \bibinfo {author} {\bibfnamefont
  {T.}~\bibnamefont {Schmitt}}, \ and\ \bibinfo {author} {\bibfnamefont
  {A.}~\bibnamefont {Keren}},\ }\href{\doibase/10.1103/PhysRevB.92.104507}
  {\bibfield  {journal} {\bibinfo  {journal} {Phys. Rev. B}\ }\textbf {\bibinfo
  {volume} {92}},\ \bibinfo {pages} {104507} (\bibinfo {year}
  {2015})}\BibitemShut {NoStop}%
\bibitem [{\citenamefont {Drachuck}\ \emph {\textit{et~al.}}(2012)\citenamefont
  {Drachuck}, \citenamefont {Shay}, \citenamefont {Bazalitsky}, \citenamefont
  {Ofer}, \citenamefont {Salman}, \citenamefont {Amato}, \citenamefont
  {Niedermayer}, \citenamefont {Wulferding}, \citenamefont {Lemmens},\ and\
  \citenamefont {Keren}}]{Drachuck2012}%
  \BibitemOpen
  \bibfield  {author} {\bibinfo {author} {\bibfnamefont {G.}~\bibnamefont
  {Drachuck}}, \bibinfo {author} {\bibfnamefont {M.}~\bibnamefont {Shay}},
  \bibinfo {author} {\bibfnamefont {G.}~\bibnamefont {Bazalitsky}}, \bibinfo
  {author} {\bibfnamefont {R.}~\bibnamefont {Ofer}}, \bibinfo {author}
  {\bibfnamefont {Z.}~\bibnamefont {Salman}}, \bibinfo {author} {\bibfnamefont
  {A.}~\bibnamefont {Amato}}, \bibinfo {author} {\bibfnamefont
  {C.}~\bibnamefont {Niedermayer}}, \bibinfo {author} {\bibfnamefont
  {D.}~\bibnamefont {Wulferding}}, \bibinfo {author} {\bibfnamefont
  {P.}~\bibnamefont {Lemmens}}, \ and\ \bibinfo {author} {\bibfnamefont
  {A.}~\bibnamefont {Keren}},\ }\href{\doibase/10.1007/s10948-012-1669-z}
  {\bibfield  {journal} {\bibinfo  {journal} {J. Supercond. Nov. Magn.}\
  }\textbf {\bibinfo {volume} {25}},\ \bibinfo {pages} {2331} (\bibinfo {year}
  {2012})}\BibitemShut {NoStop}%
\bibitem [{\citenamefont {Amit}\ and\ \citenamefont {Keren}(2010)}]{Eran2010}%
  \BibitemOpen
  \bibfield  {author} {\bibinfo {author} {\bibfnamefont {E.}~\bibnamefont
  {Amit}}\ and\ \bibinfo {author} {\bibfnamefont {A.}~\bibnamefont {Keren}},\
  }\href{\doibase/10.1103/PhysRevB.82.172509} {\bibfield  {journal} {\bibinfo
  {journal} {Phys. Rev. B}\ }\textbf {\bibinfo {volume} {82}},\ \bibinfo
  {pages} {172509} (\bibinfo {year} {2010})}\BibitemShut {NoStop}%
\bibitem [{\citenamefont {Tallon}(2014)}]{TallonPRB14}%
  \BibitemOpen
  \bibfield  {author} {\bibinfo {author} {\bibfnamefont {J.~L.}\ \bibnamefont
  {Tallon}},\ }\href{\doibase/10.1103/PhysRevB.90.214523} {\bibfield  {journal}
  {\bibinfo  {journal} {Phys. Rev. B}\ }\textbf {\bibinfo {volume} {90}},\
  \bibinfo {pages} {214523} (\bibinfo {year} {2014})}\BibitemShut {NoStop}%
\bibitem [{\citenamefont {Haverkort}\ \emph
  {\textit{et~al.}}(2010)\citenamefont {Haverkort}, \citenamefont {Hollmann},
  \citenamefont {Krug},\ and\ \citenamefont {Tanaka}}]{Haverkort2010}%
  \BibitemOpen
  \bibfield  {author} {\bibinfo {author} {\bibfnamefont {M.~W.}\ \bibnamefont
  {Haverkort}}, \bibinfo {author} {\bibfnamefont {N.}~\bibnamefont {Hollmann}},
  \bibinfo {author} {\bibfnamefont {I.~P.}\ \bibnamefont {Krug}}, \ and\
  \bibinfo {author} {\bibfnamefont {A.}~\bibnamefont {Tanaka}},\
  }\href{\doibase/10.1103/PhysRevB.82.094403} {\bibfield  {journal} {\bibinfo
  {journal} {Phys. Rev. B}\ }\textbf {\bibinfo {volume} {82}},\ \bibinfo
  {pages} {094403} (\bibinfo {year} {2010})}\BibitemShut {NoStop}%
\bibitem [{\citenamefont {Fink}\ \emph {\textit{et~al.}}(2013)\citenamefont
  {Fink}, \citenamefont {Schierle}, \citenamefont {Weschke},\ and\
  \citenamefont {Geck}}]{Fink_REXS_review}%
  \BibitemOpen
  \bibfield  {author} {\bibinfo {author} {\bibfnamefont {J.}~\bibnamefont
  {Fink}}, \bibinfo {author} {\bibfnamefont {E.}~\bibnamefont {Schierle}},
  \bibinfo {author} {\bibfnamefont {E.}~\bibnamefont {Weschke}}, \ and\
  \bibinfo {author} {\bibfnamefont {J.}~\bibnamefont {Geck}},\
  }\href{\doibase/10.1088/0034-4885/76/5/056502} {\bibfield  {journal}
  {\bibinfo  {journal} {Rep. Prog. Phys.}\ }\textbf {\bibinfo {volume} {76}},\
  \bibinfo {pages} {056502} (\bibinfo {year} {2013})}\BibitemShut {NoStop}%
\bibitem [{\citenamefont {Chmaissem}\ \emph
  {\textit{et~al.}}(2001)\citenamefont {Chmaissem}, \citenamefont {Eckstein},\
  and\ \citenamefont {Kuper}}]{Chmaissem_PRB_2001}%
  \BibitemOpen
  \bibfield  {author} {\bibinfo {author} {\bibfnamefont {O.}~\bibnamefont
  {Chmaissem}}, \bibinfo {author} {\bibfnamefont {Y.}~\bibnamefont {Eckstein}},
  \ and\ \bibinfo {author} {\bibfnamefont {C.~G.}\ \bibnamefont {Kuper}},\
  }\href{\doibase/10.1103/PhysRevB.63.174510} {\bibfield  {journal} {\bibinfo
  {journal} {Phys. Rev. B}\ }\textbf {\bibinfo {volume} {63}},\ \bibinfo
  {pages} {174510} (\bibinfo {year} {2001})}\BibitemShut {NoStop}%
\bibitem [{\citenamefont {Englisch}\ \emph {\textit{et~al.}}(2001)\citenamefont
  {Englisch}, \citenamefont {Rossner}, \citenamefont {Maletta}, \citenamefont
  {Bahrdt}, \citenamefont {Sasaki}, \citenamefont {Senf}, \citenamefont
  {Sawhney},\ and\ \citenamefont {Gudat}}]{ENGLISCH_2001_UE46}%
  \BibitemOpen
  \bibfield  {author} {\bibinfo {author} {\bibfnamefont {U.}~\bibnamefont
  {Englisch}}, \bibinfo {author} {\bibfnamefont {H.}~\bibnamefont {Rossner}},
  \bibinfo {author} {\bibfnamefont {H.}~\bibnamefont {Maletta}}, \bibinfo
  {author} {\bibfnamefont {J.}~\bibnamefont {Bahrdt}}, \bibinfo {author}
  {\bibfnamefont {S.}~\bibnamefont {Sasaki}}, \bibinfo {author} {\bibfnamefont
  {F.}~\bibnamefont {Senf}}, \bibinfo {author} {\bibfnamefont {K.}~\bibnamefont
  {Sawhney}}, \ and\ \bibinfo {author} {\bibfnamefont {W.}~\bibnamefont
  {Gudat}},\ }\href{\doibase/10.1016/S0168-9002(01)00407-7} {\bibfield
  {journal} {\bibinfo  {journal} {Nucl. Instrum. Meth. A}\ }\textbf {\bibinfo
  {volume} {467-468}},\ \bibinfo {pages} {541 } (\bibinfo {year}
  {2001})}\BibitemShut {NoStop}%
\bibitem [{\citenamefont {Achkar}\ \emph {\textit{et~al.}}(2012)\citenamefont
  {Achkar}, \citenamefont {Sutarto}, \citenamefont {Mao}, \citenamefont {He},
  \citenamefont {Frano}, \citenamefont {Blanco-Canosa}, \citenamefont
  {Le~Tacon}, \citenamefont {Ghiringhelli}, \citenamefont {Braicovich},
  \citenamefont {Minola}, \citenamefont {Moretti~Sala}, \citenamefont
  {Mazzoli}, \citenamefont {Liang}, \citenamefont {Bonn}, \citenamefont
  {Hardy}, \citenamefont {Keimer}, \citenamefont {Sawatzky},\ and\
  \citenamefont {Hawthorn}}]{Achkar_PRL_2012_Distinct_Charge_Orders}%
  \BibitemOpen
  \bibfield  {author} {\bibinfo {author} {\bibfnamefont {A.~J.}\ \bibnamefont
  {Achkar}}, \bibinfo {author} {\bibfnamefont {R.}~\bibnamefont {Sutarto}},
  \bibinfo {author} {\bibfnamefont {X.}~\bibnamefont {Mao}}, \bibinfo {author}
  {\bibfnamefont {F.}~\bibnamefont {He}}, \bibinfo {author} {\bibfnamefont
  {A.}~\bibnamefont {Frano}}, \bibinfo {author} {\bibfnamefont
  {S.}~\bibnamefont {Blanco-Canosa}}, \bibinfo {author} {\bibfnamefont
  {M.}~\bibnamefont {Le~Tacon}}, \bibinfo {author} {\bibfnamefont
  {G.}~\bibnamefont {Ghiringhelli}}, \bibinfo {author} {\bibfnamefont
  {L.}~\bibnamefont {Braicovich}}, \bibinfo {author} {\bibfnamefont
  {M.}~\bibnamefont {Minola}}, \bibinfo {author} {\bibfnamefont
  {M.}~\bibnamefont {Moretti~Sala}}, \bibinfo {author} {\bibfnamefont
  {C.}~\bibnamefont {Mazzoli}}, \bibinfo {author} {\bibfnamefont
  {R.}~\bibnamefont {Liang}}, \bibinfo {author} {\bibfnamefont {D.~A.}\
  \bibnamefont {Bonn}}, \bibinfo {author} {\bibfnamefont {W.~N.}\ \bibnamefont
  {Hardy}}, \bibinfo {author} {\bibfnamefont {B.}~\bibnamefont {Keimer}},
  \bibinfo {author} {\bibfnamefont {G.~A.}\ \bibnamefont {Sawatzky}}, \ and\
  \bibinfo {author} {\bibfnamefont {D.~G.}\ \bibnamefont {Hawthorn}},\
  }\href{\doibase/10.1103/PhysRevLett.109.167001} {\bibfield  {journal}
  {\bibinfo  {journal} {Phys. Rev. Lett.}\ }\textbf {\bibinfo {volume} {109}},\
  \bibinfo {pages} {167001} (\bibinfo {year} {2012})}\BibitemShut {NoStop}%
\bibitem [{\citenamefont {Caplan}\ \emph {\textit{et~al.}}(2015)\citenamefont
  {Caplan}, \citenamefont {Wachtel},\ and\ \citenamefont
  {Orgad}}]{Caplan_PRB_2015}%
  \BibitemOpen
  \bibfield  {author} {\bibinfo {author} {\bibfnamefont {Y.}~\bibnamefont
  {Caplan}}, \bibinfo {author} {\bibfnamefont {G.}~\bibnamefont {Wachtel}}, \
  and\ \bibinfo {author} {\bibfnamefont {D.}~\bibnamefont {Orgad}},\
  }\href{https://link.aps.org/doi/10.1103/PhysRevB.92.224504} {\bibfield
  {journal} {\bibinfo  {journal} {Phys. Rev. B}\ }\textbf {\bibinfo {volume}
  {92}},\ \bibinfo {pages} {224504} (\bibinfo {year} {2015})}\BibitemShut
  {NoStop}%
\bibitem [{\citenamefont {Caplan}\ and\ \citenamefont
  {Orgad}(2017)}]{Caplan_PRL_2017}%
  \BibitemOpen
  \bibfield  {author} {\bibinfo {author} {\bibfnamefont {Y.}~\bibnamefont
  {Caplan}}\ and\ \bibinfo {author} {\bibfnamefont {D.}~\bibnamefont {Orgad}},\
  }\href{https://link.aps.org/doi/10.1103/PhysRevLett.119.107002} {\bibfield
  {journal} {\bibinfo  {journal} {Phys. Rev. Lett.}\ }\textbf {\bibinfo
  {volume} {119}},\ \bibinfo {pages} {107002} (\bibinfo {year}
  {2017})}\BibitemShut {NoStop}%
\bibitem [{\citenamefont {da~Silva~Neto}\ \emph
  {\textit{et~al.}}(2016)\citenamefont {da~Silva~Neto}, \citenamefont {Yu},
  \citenamefont {Minola}, \citenamefont {Sutarto}, \citenamefont {Schierle},
  \citenamefont {Boschini}, \citenamefont {Zonno}, \citenamefont {Bluschke},
  \citenamefont {Higgins}, \citenamefont {Li}, \citenamefont {Yu},
  \citenamefont {Weschke}, \citenamefont {He}, \citenamefont {Le~Tacon},
  \citenamefont {Greene}, \citenamefont {Greven}, \citenamefont {Sawatzky},
  \citenamefont {Keimer},\ and\ \citenamefont {Damascelli}}]{Neto_SciAdv_2015}%
  \BibitemOpen
  \bibfield  {author} {\bibinfo {author} {\bibfnamefont {E.~H.}\ \bibnamefont
  {da~Silva~Neto}}, \bibinfo {author} {\bibfnamefont {B.}~\bibnamefont {Yu}},
  \bibinfo {author} {\bibfnamefont {M.}~\bibnamefont {Minola}}, \bibinfo
  {author} {\bibfnamefont {R.}~\bibnamefont {Sutarto}}, \bibinfo {author}
  {\bibfnamefont {E.}~\bibnamefont {Schierle}}, \bibinfo {author}
  {\bibfnamefont {F.}~\bibnamefont {Boschini}}, \bibinfo {author}
  {\bibfnamefont {M.}~\bibnamefont {Zonno}}, \bibinfo {author} {\bibfnamefont
  {M.}~\bibnamefont {Bluschke}}, \bibinfo {author} {\bibfnamefont
  {J.}~\bibnamefont {Higgins}}, \bibinfo {author} {\bibfnamefont
  {Y.}~\bibnamefont {Li}}, \bibinfo {author} {\bibfnamefont {G.}~\bibnamefont
  {Yu}}, \bibinfo {author} {\bibfnamefont {E.}~\bibnamefont {Weschke}},
  \bibinfo {author} {\bibfnamefont {F.}~\bibnamefont {He}}, \bibinfo {author}
  {\bibfnamefont {M.}~\bibnamefont {Le~Tacon}}, \bibinfo {author}
  {\bibfnamefont {R.~L.}\ \bibnamefont {Greene}}, \bibinfo {author}
  {\bibfnamefont {M.}~\bibnamefont {Greven}}, \bibinfo {author} {\bibfnamefont
  {G.~A.}\ \bibnamefont {Sawatzky}}, \bibinfo {author} {\bibfnamefont
  {B.}~\bibnamefont {Keimer}}, \ and\ \bibinfo {author} {\bibfnamefont
  {A.}~\bibnamefont {Damascelli}},\ }\href{\doibase/10.1126/sciadv.1600782}
  {\bibfield  {journal} {\bibinfo  {journal} {Sci. Adv.}\ }\textbf {\bibinfo
  {volume} {2}},\ \bibinfo {pages} {e1600782} (\bibinfo {year}
  {2016})}\BibitemShut {NoStop}%
\bibitem [{\citenamefont {Dalla~Torre}\ \emph
  {\textit{et~al.}}(2015)\citenamefont {Dalla~Torre}, \citenamefont {Benjamin},
  \citenamefont {He}, \citenamefont {Dentelski},\ and\ \citenamefont
  {Demler}}]{DallaTorre_2015_NJP}%
  \BibitemOpen
  \bibfield  {author} {\bibinfo {author} {\bibfnamefont {E.~G.}\ \bibnamefont
  {Dalla~Torre}}, \bibinfo {author} {\bibfnamefont {D.}~\bibnamefont
  {Benjamin}}, \bibinfo {author} {\bibfnamefont {Y.}~\bibnamefont {He}},
  \bibinfo {author} {\bibfnamefont {D.}~\bibnamefont {Dentelski}}, \ and\
  \bibinfo {author} {\bibfnamefont {E.}~\bibnamefont {Demler}},\
  }\href{\doibase/10.1088/1367-2630/17/2/022001} {\bibfield  {journal}
  {\bibinfo  {journal} {New J. Phys.}\ }\textbf {\bibinfo {volume} {17}},\
  \bibinfo {pages} {022001} (\bibinfo {year} {2015})}\BibitemShut {NoStop}%
\bibitem [{\citenamefont {Dalla~Torre}\ \emph
  {\textit{et~al.}}(2016)\citenamefont {Dalla~Torre}, \citenamefont {Benjamin},
  \citenamefont {He}, \citenamefont {Dentelski},\ and\ \citenamefont
  {Demler}}]{DallaTorre_2016_PRB}%
  \BibitemOpen
  \bibfield  {author} {\bibinfo {author} {\bibfnamefont {E.~G.}\ \bibnamefont
  {Dalla~Torre}}, \bibinfo {author} {\bibfnamefont {D.}~\bibnamefont
  {Benjamin}}, \bibinfo {author} {\bibfnamefont {Y.}~\bibnamefont {He}},
  \bibinfo {author} {\bibfnamefont {D.}~\bibnamefont {Dentelski}}, \ and\
  \bibinfo {author} {\bibfnamefont {E.}~\bibnamefont {Demler}},\
  }\href{\doibase/10.1103/PhysRevB.93.205117} {\bibfield  {journal} {\bibinfo
  {journal} {Phys. Rev. B}\ }\textbf {\bibinfo {volume} {93}},\ \bibinfo
  {pages} {205117} (\bibinfo {year} {2016})}\BibitemShut {NoStop}%
\bibitem [{\citenamefont {Berg}\ \emph {\textit{et~al.}}(2009)\citenamefont
  {Berg}, \citenamefont {Fradkin},\ and\ \citenamefont
  {Kivelson}}]{Berg_2009_NJP}%
  \BibitemOpen
  \bibfield  {author} {\bibinfo {author} {\bibfnamefont {E.}~\bibnamefont
  {Berg}}, \bibinfo {author} {\bibfnamefont {E.}~\bibnamefont {Fradkin}}, \
  and\ \bibinfo {author} {\bibfnamefont {S.~A.}\ \bibnamefont {Kivelson}},\
  }\href{\doibase/10.1088/1367-2630/11/11/115004} {\bibfield  {journal}
  {\bibinfo  {journal} {New J. Phys.}\ }\textbf {\bibinfo {volume} {11}},\
  \bibinfo {pages} {115004} (\bibinfo {year} {2009})}\BibitemShut {NoStop}%
\bibitem [{\citenamefont {Lee}(2014)}]{Lee2014}%
  \BibitemOpen
  \bibfield  {author} {\bibinfo {author} {\bibfnamefont {P.~A.}\ \bibnamefont
  {Lee}},\ }\href{\doibase/10.1103/PhysRevX.4.031017} {\bibfield  {journal}
  {\bibinfo  {journal} {Phys. Rev. X}\ }\textbf {\bibinfo {volume} {4}},\
  \bibinfo {pages} {031017} (\bibinfo {year} {2014})}\BibitemShut {NoStop}%
\bibitem [{\citenamefont {Blanco-Canosa}\ \emph
  {\textit{et~al.}}(2013)\citenamefont {Blanco-Canosa}, \citenamefont {Frano},
  \citenamefont {Loew}, \citenamefont {Lu}, \citenamefont {Porras},
  \citenamefont {Ghiringhelli}, \citenamefont {Minola}, \citenamefont
  {Mazzoli}, \citenamefont {Braicovich}, \citenamefont {Schierle},
  \citenamefont {Weschke}, \citenamefont {Le~Tacon},\ and\ \citenamefont
  {Keimer}}]{Santi_PRL_2013}%
  \BibitemOpen
  \bibfield  {author} {\bibinfo {author} {\bibfnamefont {S.}~\bibnamefont
  {Blanco-Canosa}}, \bibinfo {author} {\bibfnamefont {A.}~\bibnamefont
  {Frano}}, \bibinfo {author} {\bibfnamefont {T.}~\bibnamefont {Loew}},
  \bibinfo {author} {\bibfnamefont {Y.}~\bibnamefont {Lu}}, \bibinfo {author}
  {\bibfnamefont {J.}~\bibnamefont {Porras}}, \bibinfo {author} {\bibfnamefont
  {G.}~\bibnamefont {Ghiringhelli}}, \bibinfo {author} {\bibfnamefont
  {M.}~\bibnamefont {Minola}}, \bibinfo {author} {\bibfnamefont
  {C.}~\bibnamefont {Mazzoli}}, \bibinfo {author} {\bibfnamefont
  {L.}~\bibnamefont {Braicovich}}, \bibinfo {author} {\bibfnamefont
  {E.}~\bibnamefont {Schierle}}, \bibinfo {author} {\bibfnamefont
  {E.}~\bibnamefont {Weschke}}, \bibinfo {author} {\bibfnamefont
  {M.}~\bibnamefont {Le~Tacon}}, \ and\ \bibinfo {author} {\bibfnamefont
  {B.}~\bibnamefont {Keimer}},\ }\href{\doibase/10.1103/PhysRevLett.110.187001}
  {\bibfield  {journal} {\bibinfo  {journal} {Phys. Rev. Lett.}\ }\textbf
  {\bibinfo {volume} {110}},\ \bibinfo {pages} {187001} (\bibinfo {year}
  {2013})}\BibitemShut {NoStop}%
\bibitem [{\citenamefont {Sachdev}\ and\ \citenamefont
  {La~Placa}(2013)}]{Sachdev2013}%
  \BibitemOpen
  \bibfield  {author} {\bibinfo {author} {\bibfnamefont {S.}~\bibnamefont
  {Sachdev}}\ and\ \bibinfo {author} {\bibfnamefont {R.}~\bibnamefont
  {La~Placa}},\ }\href{\doibase/10.1103/PhysRevLett.111.027202} {\bibfield
  {journal} {\bibinfo  {journal} {Phys. Rev. Lett.}\ }\textbf {\bibinfo
  {volume} {111}},\ \bibinfo {pages} {027202} (\bibinfo {year}
  {2013})}\BibitemShut {NoStop}%
\bibitem [{\citenamefont {da~Silva~Neto}\ \emph
  {\textit{et~al.}}(2018)\citenamefont {da~Silva~Neto}, \citenamefont {Minola},
  \citenamefont {Yu}, \citenamefont {Tabis}, \citenamefont {Bluschke},
  \citenamefont {Unruh}, \citenamefont {Suzuki}, \citenamefont {Li},
  \citenamefont {Yu}, \citenamefont {Betto}, \citenamefont {Kummer},
  \citenamefont {Yakhou}, \citenamefont {Brookes}, \citenamefont {Le~Tacon},
  \citenamefont {Greven}, \citenamefont {Keimer},\ and\ \citenamefont
  {Damascelli}}]{Neto2018}%
  \BibitemOpen
  \bibfield  {author} {\bibinfo {author} {\bibfnamefont {E.~H.}\ \bibnamefont
  {da~Silva~Neto}}, \bibinfo {author} {\bibfnamefont {M.}~\bibnamefont
  {Minola}}, \bibinfo {author} {\bibfnamefont {B.}~\bibnamefont {Yu}}, \bibinfo
  {author} {\bibfnamefont {W.}~\bibnamefont {Tabis}}, \bibinfo {author}
  {\bibfnamefont {M.}~\bibnamefont {Bluschke}}, \bibinfo {author}
  {\bibfnamefont {D.}~\bibnamefont {Unruh}}, \bibinfo {author} {\bibfnamefont
  {H.}~\bibnamefont {Suzuki}}, \bibinfo {author} {\bibfnamefont
  {Y.}~\bibnamefont {Li}}, \bibinfo {author} {\bibfnamefont {G.}~\bibnamefont
  {Yu}}, \bibinfo {author} {\bibfnamefont {D.}~\bibnamefont {Betto}}, \bibinfo
  {author} {\bibfnamefont {K.}~\bibnamefont {Kummer}}, \bibinfo {author}
  {\bibfnamefont {F.}~\bibnamefont {Yakhou}}, \bibinfo {author} {\bibfnamefont
  {N.~B.}\ \bibnamefont {Brookes}}, \bibinfo {author} {\bibfnamefont
  {M.}~\bibnamefont {Le~Tacon}}, \bibinfo {author} {\bibfnamefont
  {M.}~\bibnamefont {Greven}}, \bibinfo {author} {\bibfnamefont
  {B.}~\bibnamefont {Keimer}}, \ and\ \bibinfo {author} {\bibfnamefont
  {A.}~\bibnamefont {Damascelli}},\ }\href{\doibase/10.1103/PhysRevB.98.161114}
  {\bibfield  {journal} {\bibinfo  {journal} {Phys. Rev. B}\ }\textbf {\bibinfo
  {volume} {98}},\ \bibinfo {pages} {161114(R)} (\bibinfo {year}
  {2018})}\BibitemShut {NoStop}%
\bibitem [{\citenamefont {Metlitski}\ and\ \citenamefont
  {Sachdev}(2010)}]{Metlitski2010}%
  \BibitemOpen
  \bibfield  {author} {\bibinfo {author} {\bibfnamefont {M.~A.}\ \bibnamefont
  {Metlitski}}\ and\ \bibinfo {author} {\bibfnamefont {S.}~\bibnamefont
  {Sachdev}},\ }\href{\doibase/10.1103/PhysRevB.82.075128} {\bibfield
  {journal} {\bibinfo  {journal} {Phys. Rev. B}\ }\textbf {\bibinfo {volume}
  {82}},\ \bibinfo {pages} {075128} (\bibinfo {year} {2010})}\BibitemShut
  {NoStop}%
\bibitem [{\citenamefont {Gerlach}\ \emph {\textit{et~al.}}(2017)\citenamefont
  {Gerlach}, \citenamefont {Schattner}, \citenamefont {Berg},\ and\
  \citenamefont {Trebst}}]{Gerlach2017}%
  \BibitemOpen
  \bibfield  {author} {\bibinfo {author} {\bibfnamefont {M.~H.}\ \bibnamefont
  {Gerlach}}, \bibinfo {author} {\bibfnamefont {Y.}~\bibnamefont {Schattner}},
  \bibinfo {author} {\bibfnamefont {E.}~\bibnamefont {Berg}}, \ and\ \bibinfo
  {author} {\bibfnamefont {S.}~\bibnamefont {Trebst}},\
  }\href{\doibase/10.1103/PhysRevB.95.035124} {\bibfield  {journal} {\bibinfo
  {journal} {Phys. Rev. B}\ }\textbf {\bibinfo {volume} {95}},\ \bibinfo
  {pages} {035124} (\bibinfo {year} {2017})}\BibitemShut {NoStop}%
\bibitem [{\citenamefont {Wang}\ \emph {\textit{et~al.}}(2018)\citenamefont
  {Wang}, \citenamefont {Wang}, \citenamefont {Schattner}, \citenamefont
  {Berg},\ and\ \citenamefont
  {Fernandes}}]{Wang_Wang_Schattner_Berg_Fernandes_PRL_2018}%
  \BibitemOpen
  \bibfield  {author} {\bibinfo {author} {\bibfnamefont {X.}~\bibnamefont
  {Wang}}, \bibinfo {author} {\bibfnamefont {Y.}~\bibnamefont {Wang}}, \bibinfo
  {author} {\bibfnamefont {Y.}~\bibnamefont {Schattner}}, \bibinfo {author}
  {\bibfnamefont {E.}~\bibnamefont {Berg}}, \ and\ \bibinfo {author}
  {\bibfnamefont {R.~M.}\ \bibnamefont {Fernandes}},\
  }\href{\doibase/10.1103/PhysRevLett.120.247002} {\bibfield  {journal}
  {\bibinfo  {journal} {Phys. Rev. Lett.}\ }\textbf {\bibinfo {volume} {120}},\
  \bibinfo {pages} {247002} (\bibinfo {year} {2018})}\BibitemShut {NoStop}%
\bibitem [{\citenamefont {Kim}\ \emph {\textit{et~al.}}(2018)\citenamefont
  {Kim}, \citenamefont {Souliou}, \citenamefont {Barber}, \citenamefont
  {Lefran{\c c}ois}, \citenamefont {Minola}, \citenamefont {Tortora},
  \citenamefont {Heid}, \citenamefont {Nandi}, \citenamefont {Borzi},
  \citenamefont {Garbarino}, \citenamefont {Bosak}, \citenamefont {Porras},
  \citenamefont {Loew}, \citenamefont {K{\"o}nig}, \citenamefont {Moll},
  \citenamefont {Mackenzie}, \citenamefont {Keimer}, \citenamefont {Hicks},\
  and\ \citenamefont {Le~Tacon}}]{Kim2019}%
  \BibitemOpen
  \bibfield  {author} {\bibinfo {author} {\bibfnamefont {H.-H.}\ \bibnamefont
  {Kim}}, \bibinfo {author} {\bibfnamefont {S.~M.}\ \bibnamefont {Souliou}},
  \bibinfo {author} {\bibfnamefont {M.~E.}\ \bibnamefont {Barber}}, \bibinfo
  {author} {\bibfnamefont {E.}~\bibnamefont {Lefran{\c c}ois}}, \bibinfo
  {author} {\bibfnamefont {M.}~\bibnamefont {Minola}}, \bibinfo {author}
  {\bibfnamefont {M.}~\bibnamefont {Tortora}}, \bibinfo {author} {\bibfnamefont
  {R.}~\bibnamefont {Heid}}, \bibinfo {author} {\bibfnamefont {N.}~\bibnamefont
  {Nandi}}, \bibinfo {author} {\bibfnamefont {R.~A.}\ \bibnamefont {Borzi}},
  \bibinfo {author} {\bibfnamefont {G.}~\bibnamefont {Garbarino}}, \bibinfo
  {author} {\bibfnamefont {A.}~\bibnamefont {Bosak}}, \bibinfo {author}
  {\bibfnamefont {J.}~\bibnamefont {Porras}}, \bibinfo {author} {\bibfnamefont
  {T.}~\bibnamefont {Loew}}, \bibinfo {author} {\bibfnamefont {M.}~\bibnamefont
  {K{\"o}nig}}, \bibinfo {author} {\bibfnamefont {P.~J.~W.}\ \bibnamefont
  {Moll}}, \bibinfo {author} {\bibfnamefont {A.~P.}\ \bibnamefont {Mackenzie}},
  \bibinfo {author} {\bibfnamefont {B.}~\bibnamefont {Keimer}}, \bibinfo
  {author} {\bibfnamefont {C.~W.}\ \bibnamefont {Hicks}}, \ and\ \bibinfo
  {author} {\bibfnamefont {M.}~\bibnamefont {Le~Tacon}},\
  }\href{\doibase/10.1126/science.aat4708} {\bibfield  {journal} {\bibinfo
  {journal} {Science}\ }\textbf {\bibinfo {volume} {362}},\ \bibinfo {pages}
  {1040} (\bibinfo {year} {2018})}\BibitemShut {NoStop}%
\bibitem [{\citenamefont {LeBoeuf}\ \emph {\textit{et~al.}}(2007)\citenamefont
  {LeBoeuf}, \citenamefont {Doiron-Leyraud}, \citenamefont {Levallois},
  \citenamefont {Daou}, \citenamefont {Bonnemaison}, \citenamefont {Hussey},
  \citenamefont {Balicas}, \citenamefont {Ramshaw}, \citenamefont {Liang},
  \citenamefont {Bonn}, \citenamefont {Hardy}, \citenamefont {Adachi},
  \citenamefont {Proust},\ and\ \citenamefont {Taillefer}}]{LeBoeuf2007}%
  \BibitemOpen
  \bibfield  {author} {\bibinfo {author} {\bibfnamefont {D.}~\bibnamefont
  {LeBoeuf}}, \bibinfo {author} {\bibfnamefont {N.}~\bibnamefont
  {Doiron-Leyraud}}, \bibinfo {author} {\bibfnamefont {J.}~\bibnamefont
  {Levallois}}, \bibinfo {author} {\bibfnamefont {R.}~\bibnamefont {Daou}},
  \bibinfo {author} {\bibfnamefont {J.-B.}\ \bibnamefont {Bonnemaison}},
  \bibinfo {author} {\bibfnamefont {N.~E.}\ \bibnamefont {Hussey}}, \bibinfo
  {author} {\bibfnamefont {L.}~\bibnamefont {Balicas}}, \bibinfo {author}
  {\bibfnamefont {B.~J.}\ \bibnamefont {Ramshaw}}, \bibinfo {author}
  {\bibfnamefont {R.}~\bibnamefont {Liang}}, \bibinfo {author} {\bibfnamefont
  {D.~A.}\ \bibnamefont {Bonn}}, \bibinfo {author} {\bibfnamefont {W.~N.}\
  \bibnamefont {Hardy}}, \bibinfo {author} {\bibfnamefont {S.}~\bibnamefont
  {Adachi}}, \bibinfo {author} {\bibfnamefont {C.}~\bibnamefont {Proust}}, \
  and\ \bibinfo {author} {\bibfnamefont {L.}~\bibnamefont {Taillefer}},\
  }\href{\doibase/10.1038/nature06332} {\bibfield  {journal} {\bibinfo
  {journal} {Nature}\ }\textbf {\bibinfo {volume} {450}},\ \bibinfo {pages}
  {533} (\bibinfo {year} {2007})}\BibitemShut {NoStop}%
\bibitem [{\citenamefont {Chang}\ \emph {\textit{et~al.}}(2010)\citenamefont
  {Chang}, \citenamefont {Daou}, \citenamefont {Proust}, \citenamefont
  {LeBoeuf}, \citenamefont {Doiron-Leyraud}, \citenamefont {Lalibert\'e},
  \citenamefont {Pingault}, \citenamefont {Ramshaw}, \citenamefont {Liang},
  \citenamefont {Bonn}, \citenamefont {Hardy}, \citenamefont {Takagi},
  \citenamefont {Antunes}, \citenamefont {Sheikin}, \citenamefont {Behnia},\
  and\ \citenamefont {Taillefer}}]{Chang2010}%
  \BibitemOpen
  \bibfield  {author} {\bibinfo {author} {\bibfnamefont {J.}~\bibnamefont
  {Chang}}, \bibinfo {author} {\bibfnamefont {R.}~\bibnamefont {Daou}},
  \bibinfo {author} {\bibfnamefont {C.}~\bibnamefont {Proust}}, \bibinfo
  {author} {\bibfnamefont {D.}~\bibnamefont {LeBoeuf}}, \bibinfo {author}
  {\bibfnamefont {N.}~\bibnamefont {Doiron-Leyraud}}, \bibinfo {author}
  {\bibfnamefont {F.}~\bibnamefont {Lalibert\'e}}, \bibinfo {author}
  {\bibfnamefont {B.}~\bibnamefont {Pingault}}, \bibinfo {author}
  {\bibfnamefont {B.~J.}\ \bibnamefont {Ramshaw}}, \bibinfo {author}
  {\bibfnamefont {R.}~\bibnamefont {Liang}}, \bibinfo {author} {\bibfnamefont
  {D.~A.}\ \bibnamefont {Bonn}}, \bibinfo {author} {\bibfnamefont {W.~N.}\
  \bibnamefont {Hardy}}, \bibinfo {author} {\bibfnamefont {H.}~\bibnamefont
  {Takagi}}, \bibinfo {author} {\bibfnamefont {A.~B.}\ \bibnamefont {Antunes}},
  \bibinfo {author} {\bibfnamefont {I.}~\bibnamefont {Sheikin}}, \bibinfo
  {author} {\bibfnamefont {K.}~\bibnamefont {Behnia}}, \ and\ \bibinfo {author}
  {\bibfnamefont {L.}~\bibnamefont {Taillefer}},\
  }\href{\doibase/10.1103/PhysRevLett.104.057005} {\bibfield  {journal}
  {\bibinfo  {journal} {Phys. Rev. Lett.}\ }\textbf {\bibinfo {volume} {104}},\
  \bibinfo {pages} {057005} (\bibinfo {year} {2010})}\BibitemShut {NoStop}%
\bibitem [{\citenamefont {Doiron-Leyraud}\ \emph
  {\textit{et~al.}}(2013)\citenamefont {Doiron-Leyraud}, \citenamefont
  {Lepault}, \citenamefont {Cyr-Choini\`ere}, \citenamefont {Vignolle},
  \citenamefont {Grissonnanche}, \citenamefont {Lalibert\'e}, \citenamefont
  {Chang}, \citenamefont {Bari\ifmmode \check{s}\else
  \v{s}\fi{}i\ifmmode~\acute{c}\else \'{c}\fi{}}, \citenamefont {Chan},
  \citenamefont {Ji}, \citenamefont {Zhao}, \citenamefont {Li}, \citenamefont
  {Greven}, \citenamefont {Proust},\ and\ \citenamefont
  {Taillefer}}]{Doiron-Leyraud_2013_Hg1201}%
  \BibitemOpen
  \bibfield  {author} {\bibinfo {author} {\bibfnamefont {N.}~\bibnamefont
  {Doiron-Leyraud}}, \bibinfo {author} {\bibfnamefont {S.}~\bibnamefont
  {Lepault}}, \bibinfo {author} {\bibfnamefont {O.}~\bibnamefont
  {Cyr-Choini\`ere}}, \bibinfo {author} {\bibfnamefont {B.}~\bibnamefont
  {Vignolle}}, \bibinfo {author} {\bibfnamefont {G.}~\bibnamefont
  {Grissonnanche}}, \bibinfo {author} {\bibfnamefont {F.}~\bibnamefont
  {Lalibert\'e}}, \bibinfo {author} {\bibfnamefont {J.}~\bibnamefont {Chang}},
  \bibinfo {author} {\bibfnamefont {N.}~\bibnamefont {Bari\ifmmode
  \check{s}\else \v{s}\fi{}i\ifmmode~\acute{c}\else \'{c}\fi{}}}, \bibinfo
  {author} {\bibfnamefont {M.~K.}\ \bibnamefont {Chan}}, \bibinfo {author}
  {\bibfnamefont {L.}~\bibnamefont {Ji}}, \bibinfo {author} {\bibfnamefont
  {X.}~\bibnamefont {Zhao}}, \bibinfo {author} {\bibfnamefont {Y.}~\bibnamefont
  {Li}}, \bibinfo {author} {\bibfnamefont {M.}~\bibnamefont {Greven}}, \bibinfo
  {author} {\bibfnamefont {C.}~\bibnamefont {Proust}}, \ and\ \bibinfo {author}
  {\bibfnamefont {L.}~\bibnamefont {Taillefer}},\
  }\href{\doibase/10.1103/PhysRevX.3.021019} {\bibfield  {journal} {\bibinfo
  {journal} {Phys. Rev. X}\ }\textbf {\bibinfo {volume} {3}},\ \bibinfo {pages}
  {021019} (\bibinfo {year} {2013})}\BibitemShut {NoStop}%
\bibitem [{\citenamefont {Doiron-Leyraud}\ \emph
  {\textit{et~al.}}(2007)\citenamefont {Doiron-Leyraud}, \citenamefont
  {Proust}, \citenamefont {LeBoeuf}, \citenamefont {Levallois}, \citenamefont
  {Bonnemaison}, \citenamefont {Liang}, \citenamefont {Bonn}, \citenamefont
  {Hardy},\ and\ \citenamefont {Taillefer}}]{Doiron-Leyraud2007}%
  \BibitemOpen
  \bibfield  {author} {\bibinfo {author} {\bibfnamefont {N.}~\bibnamefont
  {Doiron-Leyraud}}, \bibinfo {author} {\bibfnamefont {C.}~\bibnamefont
  {Proust}}, \bibinfo {author} {\bibfnamefont {D.}~\bibnamefont {LeBoeuf}},
  \bibinfo {author} {\bibfnamefont {J.}~\bibnamefont {Levallois}}, \bibinfo
  {author} {\bibfnamefont {J.-B.}\ \bibnamefont {Bonnemaison}}, \bibinfo
  {author} {\bibfnamefont {R.}~\bibnamefont {Liang}}, \bibinfo {author}
  {\bibfnamefont {D.~A.}\ \bibnamefont {Bonn}}, \bibinfo {author}
  {\bibfnamefont {W.~N.}\ \bibnamefont {Hardy}}, \ and\ \bibinfo {author}
  {\bibfnamefont {L.}~\bibnamefont {Taillefer}},\
  }\href{\doibase/10.1038/nature05872} {\bibfield  {journal} {\bibinfo
  {journal} {Nature}\ }\textbf {\bibinfo {volume} {447}},\ \bibinfo {pages}
  {565} (\bibinfo {year} {2007})}\BibitemShut {NoStop}%
\bibitem [{\citenamefont {Yelland}\ \emph {\textit{et~al.}}(2008)\citenamefont
  {Yelland}, \citenamefont {Singleton}, \citenamefont {Mielke}, \citenamefont
  {Harrison}, \citenamefont {Balakirev}, \citenamefont {Dabrowski},\ and\
  \citenamefont {Cooper}}]{Yelland2008}%
  \BibitemOpen
  \bibfield  {author} {\bibinfo {author} {\bibfnamefont {E.~A.}\ \bibnamefont
  {Yelland}}, \bibinfo {author} {\bibfnamefont {J.}~\bibnamefont {Singleton}},
  \bibinfo {author} {\bibfnamefont {C.~H.}\ \bibnamefont {Mielke}}, \bibinfo
  {author} {\bibfnamefont {N.}~\bibnamefont {Harrison}}, \bibinfo {author}
  {\bibfnamefont {F.~F.}\ \bibnamefont {Balakirev}}, \bibinfo {author}
  {\bibfnamefont {B.}~\bibnamefont {Dabrowski}}, \ and\ \bibinfo {author}
  {\bibfnamefont {J.~R.}\ \bibnamefont {Cooper}},\
  }\href{\doibase/10.1103/PhysRevLett.100.047003} {\bibfield  {journal}
  {\bibinfo  {journal} {Phys. Rev. Lett.}\ }\textbf {\bibinfo {volume} {100}},\
  \bibinfo {pages} {047003} (\bibinfo {year} {2008})}\BibitemShut {NoStop}%
\bibitem [{\citenamefont {Bari\v{s}i\'c}\ \emph
  {\textit{et~al.}}(2013)\citenamefont {Bari\v{s}i\'c}, \citenamefont {Badoux},
  \citenamefont {Chan}, \citenamefont {Dorow}, \citenamefont {Tabis},
  \citenamefont {Vignolle}, \citenamefont {Yu}, \citenamefont {B{\'e}ard},
  \citenamefont {Zhao}, \citenamefont {Proust},\ and\ \citenamefont
  {Greven}}]{Barisic2013}%
  \BibitemOpen
  \bibfield  {author} {\bibinfo {author} {\bibfnamefont {N.}~\bibnamefont
  {Bari\v{s}i\'c}}, \bibinfo {author} {\bibfnamefont {S.}~\bibnamefont
  {Badoux}}, \bibinfo {author} {\bibfnamefont {M.~K.}\ \bibnamefont {Chan}},
  \bibinfo {author} {\bibfnamefont {C.}~\bibnamefont {Dorow}}, \bibinfo
  {author} {\bibfnamefont {W.}~\bibnamefont {Tabis}}, \bibinfo {author}
  {\bibfnamefont {B.}~\bibnamefont {Vignolle}}, \bibinfo {author}
  {\bibfnamefont {G.}~\bibnamefont {Yu}}, \bibinfo {author} {\bibfnamefont
  {J.}~\bibnamefont {B{\'e}ard}}, \bibinfo {author} {\bibfnamefont
  {X.}~\bibnamefont {Zhao}}, \bibinfo {author} {\bibfnamefont {C.}~\bibnamefont
  {Proust}}, \ and\ \bibinfo {author} {\bibfnamefont {M.}~\bibnamefont
  {Greven}},\ }\href@noop {} {\bibfield  {journal} {\bibinfo  {journal} {Nat.
  Phys.}\ }\textbf {\bibinfo {volume} {9}},\ \bibinfo {pages} {761} (\bibinfo
  {year} {2013})}\BibitemShut {NoStop}%
\bibitem [{\citenamefont {Wu}\ \emph {\textit{et~al.}}(2011)\citenamefont {Wu},
  \citenamefont {Mayaffre}, \citenamefont {Kramer}, \citenamefont {Horvatic},
  \citenamefont {Berthier}, \citenamefont {Hardy}, \citenamefont {Liang},
  \citenamefont {Bonn},\ and\ \citenamefont {Julien}}]{Wu2011}%
  \BibitemOpen
  \bibfield  {author} {\bibinfo {author} {\bibfnamefont {T.}~\bibnamefont
  {Wu}}, \bibinfo {author} {\bibfnamefont {H.}~\bibnamefont {Mayaffre}},
  \bibinfo {author} {\bibfnamefont {S.}~\bibnamefont {Kramer}}, \bibinfo
  {author} {\bibfnamefont {M.}~\bibnamefont {Horvatic}}, \bibinfo {author}
  {\bibfnamefont {C.}~\bibnamefont {Berthier}}, \bibinfo {author}
  {\bibfnamefont {W.~N.}\ \bibnamefont {Hardy}}, \bibinfo {author}
  {\bibfnamefont {R.}~\bibnamefont {Liang}}, \bibinfo {author} {\bibfnamefont
  {D.~A.}\ \bibnamefont {Bonn}}, \ and\ \bibinfo {author} {\bibfnamefont
  {M.-H.}\ \bibnamefont {Julien}},\ }\href{\doibase/10.1038/nature10345}
  {\bibfield  {journal} {\bibinfo  {journal} {Nature}\ }\textbf {\bibinfo
  {volume} {477}},\ \bibinfo {pages} {191} (\bibinfo {year}
  {2011})}\BibitemShut {NoStop}%
\bibitem [{\citenamefont {Gerber}\ \emph {\textit{et~al.}}(2015)\citenamefont
  {Gerber}, \citenamefont {Jang}, \citenamefont {Nojiri}, \citenamefont
  {Matsuzawa}, \citenamefont {Yasumura}, \citenamefont {Bonn}, \citenamefont
  {Liang}, \citenamefont {Hardy}, \citenamefont {Islam}, \citenamefont {Mehta},
  \citenamefont {Song}, \citenamefont {Sikorski}, \citenamefont {Stefanescu},
  \citenamefont {Feng}, \citenamefont {Kivelson}, \citenamefont {Devereaux},
  \citenamefont {Shen}, \citenamefont {Kao}, \citenamefont {Lee}, \citenamefont
  {Zhu},\ and\ \citenamefont {Lee}}]{Gerber949}%
  \BibitemOpen
  \bibfield  {author} {\bibinfo {author} {\bibfnamefont {S.}~\bibnamefont
  {Gerber}}, \bibinfo {author} {\bibfnamefont {H.}~\bibnamefont {Jang}},
  \bibinfo {author} {\bibfnamefont {H.}~\bibnamefont {Nojiri}}, \bibinfo
  {author} {\bibfnamefont {S.}~\bibnamefont {Matsuzawa}}, \bibinfo {author}
  {\bibfnamefont {H.}~\bibnamefont {Yasumura}}, \bibinfo {author}
  {\bibfnamefont {D.~A.}\ \bibnamefont {Bonn}}, \bibinfo {author}
  {\bibfnamefont {R.}~\bibnamefont {Liang}}, \bibinfo {author} {\bibfnamefont
  {W.~N.}\ \bibnamefont {Hardy}}, \bibinfo {author} {\bibfnamefont
  {Z.}~\bibnamefont {Islam}}, \bibinfo {author} {\bibfnamefont
  {A.}~\bibnamefont {Mehta}}, \bibinfo {author} {\bibfnamefont
  {S.}~\bibnamefont {Song}}, \bibinfo {author} {\bibfnamefont {M.}~\bibnamefont
  {Sikorski}}, \bibinfo {author} {\bibfnamefont {D.}~\bibnamefont
  {Stefanescu}}, \bibinfo {author} {\bibfnamefont {Y.}~\bibnamefont {Feng}},
  \bibinfo {author} {\bibfnamefont {S.~A.}\ \bibnamefont {Kivelson}}, \bibinfo
  {author} {\bibfnamefont {T.~P.}\ \bibnamefont {Devereaux}}, \bibinfo {author}
  {\bibfnamefont {Z.-X.}\ \bibnamefont {Shen}}, \bibinfo {author}
  {\bibfnamefont {C.-C.}\ \bibnamefont {Kao}}, \bibinfo {author} {\bibfnamefont
  {W.-S.}\ \bibnamefont {Lee}}, \bibinfo {author} {\bibfnamefont
  {D.}~\bibnamefont {Zhu}}, \ and\ \bibinfo {author} {\bibfnamefont {J.-S.}\
  \bibnamefont {Lee}},\ }\href{\doibase/10.1126/science.aac6257} {\bibfield
  {journal} {\bibinfo  {journal} {Science}\ }\textbf {\bibinfo {volume}
  {350}},\ \bibinfo {pages} {949} (\bibinfo {year} {2015})}\BibitemShut
  {NoStop}%
\bibitem [{\citenamefont {Chang}\ \emph {\textit{et~al.}}(2016)\citenamefont
  {Chang}, \citenamefont {Blackburn}, \citenamefont {Ivashko}, \citenamefont
  {Holmes}, \citenamefont {Christensen}, \citenamefont {H{\"u}cker},
  \citenamefont {Liang}, \citenamefont {Bonn}, \citenamefont {Hardy},
  \citenamefont {R{\"u}tt}, \citenamefont {Zimmermann}, \citenamefont
  {Forgan},\ and\ \citenamefont {Hayden}}]{Chang2016}%
  \BibitemOpen
  \bibfield  {author} {\bibinfo {author} {\bibfnamefont {J.}~\bibnamefont
  {Chang}}, \bibinfo {author} {\bibfnamefont {E.}~\bibnamefont {Blackburn}},
  \bibinfo {author} {\bibfnamefont {O.}~\bibnamefont {Ivashko}}, \bibinfo
  {author} {\bibfnamefont {A.~T.}\ \bibnamefont {Holmes}}, \bibinfo {author}
  {\bibfnamefont {N.~B.}\ \bibnamefont {Christensen}}, \bibinfo {author}
  {\bibfnamefont {M.}~\bibnamefont {H{\"u}cker}}, \bibinfo {author}
  {\bibfnamefont {R.}~\bibnamefont {Liang}}, \bibinfo {author} {\bibfnamefont
  {D.~A.}\ \bibnamefont {Bonn}}, \bibinfo {author} {\bibfnamefont {W.~N.}\
  \bibnamefont {Hardy}}, \bibinfo {author} {\bibfnamefont {U.}~\bibnamefont
  {R{\"u}tt}}, \bibinfo {author} {\bibfnamefont {M.~v.}\ \bibnamefont
  {Zimmermann}}, \bibinfo {author} {\bibfnamefont {E.~M.}\ \bibnamefont
  {Forgan}}, \ and\ \bibinfo {author} {\bibfnamefont {S.~M.}\ \bibnamefont
  {Hayden}},\ }\href{https://www.nature.com/articles/ncomms11494} {\bibfield
  {journal} {\bibinfo  {journal} {Nat. Commun.}\ }\textbf {\bibinfo {volume}
  {7}},\ \bibinfo {pages} {11494} (\bibinfo {year} {2016})}\BibitemShut
  {NoStop}%
\bibitem [{\citenamefont {Lalibert{\'e}}\ \emph
  {\textit{et~al.}}(2018)\citenamefont {Lalibert{\'e}}, \citenamefont
  {Frachet}, \citenamefont {Benhabib}, \citenamefont {Borgnic}, \citenamefont
  {Loew}, \citenamefont {Porras}, \citenamefont {Le~Tacon}, \citenamefont
  {Keimer}, \citenamefont {Wiedmann}, \citenamefont {Proust},\ and\
  \citenamefont {LeBoeuf}}]{Laliberte2018}%
  \BibitemOpen
  \bibfield  {author} {\bibinfo {author} {\bibfnamefont {F.}~\bibnamefont
  {Lalibert{\'e}}}, \bibinfo {author} {\bibfnamefont {M.}~\bibnamefont
  {Frachet}}, \bibinfo {author} {\bibfnamefont {S.}~\bibnamefont {Benhabib}},
  \bibinfo {author} {\bibfnamefont {B.}~\bibnamefont {Borgnic}}, \bibinfo
  {author} {\bibfnamefont {T.}~\bibnamefont {Loew}}, \bibinfo {author}
  {\bibfnamefont {J.}~\bibnamefont {Porras}}, \bibinfo {author} {\bibfnamefont
  {M.}~\bibnamefont {Le~Tacon}}, \bibinfo {author} {\bibfnamefont
  {B.}~\bibnamefont {Keimer}}, \bibinfo {author} {\bibfnamefont
  {S.}~\bibnamefont {Wiedmann}}, \bibinfo {author} {\bibfnamefont
  {C.}~\bibnamefont {Proust}}, \ and\ \bibinfo {author} {\bibfnamefont
  {D.}~\bibnamefont {LeBoeuf}},\ }\href{\doibase/10.1038/s41535-018-0084-5}
  {\bibfield  {journal} {\bibinfo  {journal} {NPJ Quant. Mater.}\ }\textbf
  {\bibinfo {volume} {3}},\ \bibinfo {pages} {11} (\bibinfo {year}
  {2018})}\BibitemShut {NoStop}%
\end{thebibliography}

%

\end{document}